\documentclass[12pt, draftclsnofoot, onecolumn]{IEEEtran}
\usepackage{cite}
\usepackage{url}
\usepackage{xcolor}
\usepackage{amsmath,amssymb,bm,amsfonts,amsthm,mathrsfs}
\usepackage{caption}
\usepackage{setspace}
\usepackage{bbm,multicol,multirow,makecell}
\usepackage{algorithm,algorithmic}
\usepackage{array,epsfig}
\usepackage{graphicx,subfig}
\usepackage{stfloats}
\usepackage{lipsum}
\usepackage{setspace}

\usepackage[T1]{fontenc}
\theoremstyle{definition} % non-italic text in the following environments
\newtheorem{theorem}{Theorem}

\newtheorem{remark}{Remark}
\newtheorem{lemma}{Lemma}

\newtheorem{corollary}{Corollary}

\newtheorem{proposition}{Proposition}
\newtheorem{definition}{Definition}

\allowdisplaybreaks[4]
\begin{document}

\title{Cluster-Free NOMA Communications Towards Next Generation Multiple Access}
\author{Xiaoxia Xu,
Yuanwei Liu, ~\IEEEmembership{Senior Member,~IEEE,}
Xidong Mu, ~\IEEEmembership{Member,~IEEE,}
Qimei Chen, ~\IEEEmembership{Member,~IEEE,}
and Zhiguo Ding, ~\IEEEmembership{Fellow,~IEEE}
\thanks{X. Xu and Q. Chen are with the School of Electronic Information, Wuhan University, Wuhan, 430072, China (e-mail: \{xiaoxiaxu, chenqimei\}@whu.edu.cn).}
\thanks{Y. Liu and X. Mu are with the School of Electronic Engineering and Computer Science, Queen Mary University of
London, London E1 4NS, U.K. (email: \{yuanwei.liu,xidong.mu\}@qmul.ac.uk).}
%\thanks{X. Mu is with School of Artificial Intelligence, Beijing University of Posts and Telecommunications, Beijing, 100876, China (email: muxidong@bupt.edu.cn).}
\thanks{Z. Ding is with the School of Electrical and Electronic Engineering, The University of Manchester, Manchester M13 9PL, U.K. (email: zhiguo.ding@manchester.ac.uk).}
}
\maketitle

%\author{
%\IEEEauthorblockN{ Yuanwei~Liu\IEEEauthorrefmark{1}, Zhijin~Qin\IEEEauthorrefmark{1}, Maged Elkashlan\IEEEauthorrefmark{1}, and  Yue~Gao\IEEEauthorrefmark{1}\\} \IEEEauthorblockA{
%\IEEEauthorrefmark{1} Queen Mary University of London, London, UK\\
%%\IEEEauthorrefmark{2} Lancaster University, Lancaster, UK\\
% } }

\maketitle
\vspace{-1.5cm}
\begin{abstract}
\vspace{-0.2cm}
A generalized downlink multi-antenna non-orthogonal multiple access (NOMA) transmission framework is proposed with the novel concept of cluster-free successive interference cancellation (SIC). In contrast to conventional NOMA approaches, where SIC is successively carried out within the same cluster, the key idea is that the SIC can be flexibly implemented between any arbitrary users to achieve efficient interference elimination. Based on the proposed framework, a sum rate maximization problem is formulated for jointly optimizing the transmit beamforming and the SIC operations between users, subject to the SIC decoding conditions and users' minimal data rate requirements. To tackle this highly-coupled mixed-integer nonlinear programming problem, an alternating direction method of multipliers-successive convex approximation (ADMM-SCA) algorithm is developed. The original problem is first reformulated into a tractable biconvex augmented Lagrangian (AL) problem by handling the non-convex terms via SCA. Then, this AL problem is decomposed into two subproblems that are iteratively solved by the ADMM to obtain the stationary solution. Furthermore, to reduce the computational complexity and alleviate the parameter initialization sensitivity of ADMM-SCA, a Matching-SCA algorithm is proposed. The intractable binary SIC operations are solved through an extended many-to-many matching, which is jointly combined with an SCA process to optimize the transmit beamforming. The proposed Matching-SCA can converge to an enhanced exchange-stable matching that guarantees the local optimality. Numerical results demonstrate that: i) the proposed Matching-SCA algorithm achieves comparable performance and a faster convergence compared to ADMM-SCA; ii) the proposed generalized framework realizes  \textit{scenario-adaptive} communications and outperforms traditional multi-antenna NOMA approaches in various communication regimes.
\end{abstract}
%\vspace{-0.5em}
\begin{IEEEkeywords}
\vspace{-0.5em}
{N}ext-generation multiple access (NGMA), non-orthogonal multiple access (NOMA), multiple antennas, successive interference cancellation (SIC).
\end{IEEEkeywords}
\vspace{-0.8em}

\section{Introduction}
Wireless communications are currently undergoing an unprecedented revolution. It is predicted by Cisco that the number of wireless-enabled devices will increase to more than 40 billion by 2023 \cite{Cisco}. Furthermore, the types of future wireless-enabled devices will vary from smart phones to connected cars, wearables, sensors, collaborative robots, and so on. Due to the explosive demands of wireless traffics and the emergence of various innovative wireless applications, next-generation wireless network, also referred to as the sixth generation (6G), is evolving towards a new era of the Internet of Everything (IoE) \cite{6GIoE}. Driven by this exciting vision, 6G is expected to embrace broadband-hungry transmissions, pervasive access, and extremely massive connectivity in diverse and heterogeneous communication scenarios \cite{6GRoadmap}. To meet these challenges, the realization of 6G requires a fully integration and a seamless convergence of different multiple access technologies, namely next generation multiple access (NGMA) \cite{EvolNOMA}.
As a promising multiple access technology, power-domain non-orthogonal multiple access (NOMA)\footnote{For the sake of expression, we refer to the power-domain NOMA as NOMA in this paper.} \cite{PDNOMA_2017,NOMA5G_2017} has become an indispensable component of NGMA. 
By exploiting the signal superposition at transmitters and the successive interference cancellation (SIC) at receivers, NOMA enables users served by the same time/frequency/space/code resource block to be further multiplexed and distinguished in the power domain.
Hence, it can dramatically enhance the network capacity and user connections, as well as reducing the outage probability \cite{Fairness_NOMA_2018}. 

On the road from NOMA to NGMA, the integration of NOMA and multiple-antenna multiple-output multiple-input (MIMO) technologies has been regarded as one key aspect \cite{MIMONOMA_Huang_2018}.
On the one hand, multiple-antenna technologies can enable spatial-domain multi access (SDMA) and provide additional spatial degrees of freedom (DoFs) to assist NOMA communications. 
On the other hand, NOMA opens up new dimensions and opportunities for resource reuse, which is capable of increasing the affordable traffic loadings of multiple-antenna communication systems \cite{MassiveAccess}.
Therefore, multi-antenna NOMA provides a promising way to significantly improve spectral efficiency and connection density for next-generation wireless systems \cite{MIMONOMA_Huang_2018,MIMONOMA_2016}.

\vspace{-0.8em}
\subsection{Prior Works}
\vspace{-0.4em}
In the past few years, extensive literatures have been devoted to the development of multi-antenna NOMA systems.
Existing multi-antenna NOMA systems can be loosely classified into two categories, namely beamformer-based NOMA and cluster-based NOMA, which differ in the strategies of both multi-antenna beamforming and SIC operation designs \cite{MIMONOMA_2018}.
\subsubsection{Studies on beamformer-based NOMA}
Beamformer-based NOMA \cite{BBNOMA_2016_Hanif,BBNOMA_2015,BBNOMA_2016_Chen,BBNOMA_Chen} directly serves different users via distinct beamforming vectors, whose beamforming strategy is similar to conventional multiple-antenna communication systems.
Meanwhile, by carrying out SIC between the multiplexed users, the spatial interference that cannot be effectively mitigated by beamforming can be further suppressed leveraging NOMA.
Based on a minorization-maximization algorithm, the authors of \cite{BBNOMA_2016_Hanif} optimized the beamformer to maximize the sum rate for a multi-user downlink multiple-input single-output NOMA (MISO-NOMA) system.
Simulation results signified that beamformer-based NOMA outperforms the traditional multi-antenna communication systems in the severely overloaded systems, where the transmit antenna number is much larger than the user number.
Additionally, the authors of \cite{BBNOMA_2015} investigated the optimal power allocation in a two-user downlink MIMO-NOMA system, which can achieve the capacity region of the MIMO broadcast channel under the derived channel state information (CSI) condition.
The authors of \cite{BBNOMA_2016_Chen} derived the condition of quasi-degraded channels, based on which a low-complexity precoding scheme was proposed for multi-user MISO-NOMA transmissions to approach the rate region of the dirty paper coding.
By considering both perfect and imperfect CSI cases, the authors of \cite{BBNOMA_Chen} further proposed low-complexity beamforming and user selection schemes to improve the sum rate and the outage probability of beamformer-based NOMA systems.

\subsubsection{Studies on cluster-based NOMA}
Different from the beamformer-based NOMA, cluster-based NOMA \cite{MIMONOMA_2016,CBNOMA_OMA,EE_CBNOMA_2019,CBNOMA_2019_Hu} typically divides the highly channel correlated users into the same cluster, where each cluster shares the same beamforming vector.
While the inter-cluster interference is mitigated via beamforming, the intra-cluster interference is suppressed by carrying out SIC within each cluster \cite{MIMONOMA_2016}.
In \cite{CBNOMA_OMA}, the authors analysed the performance of the cluster-based NOMA system, which analytically demonstrated the superiority of cluster-based NOMA over the MIMO and orthogonal multiple-access (MIMO-OMA) system in terms of both sum channel capacity and ergodic sum capacity.
The authors of \cite{EE_CBNOMA_2019} investigated an uplink millimeter-wave (mmWave) massive MIMO-NOMA system with hybrid anglog-digital beamforming, where user clustering was obtained by considering both users' channel correlations and gain difference, and the power allocation is designed to maximize the energy efficiency. 
In \cite{CBNOMA_2019_Hu}, the authors proposed a two-stage cluster grouping algorithm for an angle-domain mmWave MIMO-NOMA system, and investigated the max-min power control to enhance user fairness. 
Considering multi-cell MISO-NOMA system, the authors of \cite{Multicell_CBNOMA_2020} proposed a distributed user grouping, beamforming and power control algorithm for power consumption minimization.
Furthermore, the author of \cite{NOMABFDesign_Ding} investigated two different NOMA beamfoming schemes, where the NOMA user shares the spatial beam with legacy SDMA users or exploits a dedicated beam. The optimal solution for both schemes are analyzed, and the studies showed that sharing spatial beam can significantly reduce the computational complexity at the expense of a slight performance loss.

\vspace{-1em}
\subsection{Motivations and Contribution}
Note that SIC plays an important role in NOMA and the design of SIC operations between users is crucial for the eventual performance achieved by NOMA.
As discussed above, current multi-antenna NOMA approaches \cite{MIMONOMA_Huang_2018,MIMONOMA_2016,MIMONOMA_2018,BBNOMA_2016_Hanif,BBNOMA_2015,BBNOMA_2016_Chen,BBNOMA_Chen,CBNOMA_OMA,Multicell_CBNOMA_2020,EE_CBNOMA_2019,CBNOMA_2019_Hu} generally assume that the SIC is sequentially carried out within the same cluster, namely cluster-specific SIC, thus leading to both benefits and drawbacks.
To be more specific, on the one hand, beamformer-based NOMA assigns all users to a single cluster, which is shown to be capable of achieving the same performance as the dirty paper coding scheme in some specific scenarios \cite{BBNOMA_2016_Chen}. However, given the sequential nature of cluster-specific SIC, users in higher SIC decoding orders have to implement a large number of SIC operations before decoding their own signals, thus leading to a high system complexity.
Moreover, beamformer-based NOMA also encounters the \emph{SIC overuse} issue, especially when users' channels are low-correlated \cite{EvolNOMA}.
This is because the SIC decoding conditions can impose undesired spatial interference to low channel-correlated users, even if this interference could have been eliminated via the spatial multiplexing.
On the other hand, cluster-based NOMA partially alleviates the \emph{SIC overuse} issue by dividing users into different clusters, where the inter-cluster and intra-cluster interference can be mitigated via spatially separated beamforming and SIC, respectively. Therefore, it can support a large number of users with a moderate SIC complexity. However, cluster-based NOMA relies on the assumption that the users in the same cluster have high channel correlations while the users of different clusters experiencing low channel correlations, which may not always hold due to the randomness of wireless channels.

It can be observed that both beamformer-based NOMA and cluster-based NOMA are \textit{scenario-centric}, whose effectiveness depends on specific scenarios, and thus cannot meet the heterogeneous scenario challenges for next-generation wireless networks.
Against this background and to pave the way to NGMA, this paper proposes a novel generalized downlink multi-antenna NOMA transmission framework with the concept of cluster-free SIC.
It enables SIC to be flexibly implemented over any arbitrary non-orthogonal users to achieve efficient interference elimination, thus breaking the constraints of the existing cluster-specific multi-antenna NOMA approaches.
Mathematically, it provides a generalized modelling, which not only unifies the existing approaches but also provides more flexible transmission options, thus overcoming the shortcoming of existing approaches.
This enables a paradigm of \textit{scenario-adaptive} multi-antenna NOMA for NGMA.
The contributions of this paper can be summarized as follows.
\begin{itemize}
  \item We propose a novel generalized downlink multi-antenna NOMA transmission framework with the concept of cluster-free SIC, which enables flexible SIC operations between users to  facilitate efficient interference elimination. The proposed framework can overcome shortcomings of traditional methods and empower a \textit{scenario-adaptive} multi-antenna NOMA paradigm.
      We formulate a sum rate maximization problem for jointly optimizing the transmit beamforming and the SIC operations subject to SIC decoding conditions and users' data rate constraints.
  \item We develop an alternating direction method of multipliers-successive convex approximation (ADMM-SCA) algorithm to tackle the formulated mixed-integer nonlinear programming (MINLP) problem, which is highly coupled and non-convex. The original problem is first reformulated into a tractable augmented Lagrangian (AL) problem, where the non-convex terms are handled by invoking the SCA method. The obtained biconvex AL problem is then decomposed into two convex subproblems, which are iteratively solved by ADMM to obtain a stationary solution.
  \item We propose a Matching-SCA algorithm to further reduce the computational complexity and overcome the parameter initialization sensitivity of ADMM-SCA.
    The SIC operations between users are modelled as a two-sided many-to-many matching with externality.
  Then, we extend the conventional swap-based matching to efficiently solve the SIC operation problem, while employing the SCA to jointly optimize the corresponding transmit beamforming. The proposed Matching-SCA can converge to an enhanced exchange-stable matching, which guarantees the local optimality.
  \item Numerical results demonstrate that the proposed Matching-SCA algorithm results in comparable performance and a faster convergence compared to the ADMM-SCA algorithm, especially in the overloaded regime.
      It is also shown that the proposed generalized multi-antenna NOMA framework is capable of achieving efficient SIC operations and \textit{scenario-adaptive} communications, which outperforms traditional transmission schemes regardless of system loadings and users' channel correlations.
\end{itemize}

\vspace{-1.2em}
\subsection{Organization and Notation}
\vspace{-0.4em}
The rest of this paper is organized as follows.
Section II presents the generalized downlink multi-antenna NOMA transmission framework and formulates the sum rate maximization problem.
In Section III, an ADMM-SCA algorithm is developed for solving the formulated joint optimization problem.
Furthermore, a low-complexity and fast-convergent Matching-SCA algorithm is proposed in Section IV by extending the conventional many-to-many matching theory.
Section V presents numerical results to demonstrate efficiencies of the proposed algorithms,
and Section VI finally concludes the paper.

\textit{Notation}: Vectors and matrices are denoted by bold-face %lower-case and upper-case
letters.
$\mathscr{R}\left(x\right)$ represents the real part of a complex variable $x$.
$\|\mathbf{x}\|$ denotes the Euclidean norm of a vector $\mathbf{x}$.
$\mathbf{x}^{T}$ and $\mathbf{x}^{H}$ denote the transpose and Hermitian conjugate of vector $\mathbf{x}$, respectively.
$\mathbf{I}_{N\times N}$ indicates an identity matrix of size $N$.
$\mathbf{1}_{M\times N}$ denotes an $M\times N$ all-ones matrix.

\section{System Model and Problem Formulation}
\subsection{A Generalized Cluster-Free Multi-Antenna NOMA Framework}

\begin{figure}[h]
  \centering
  \includegraphics[width=0.9\textwidth]{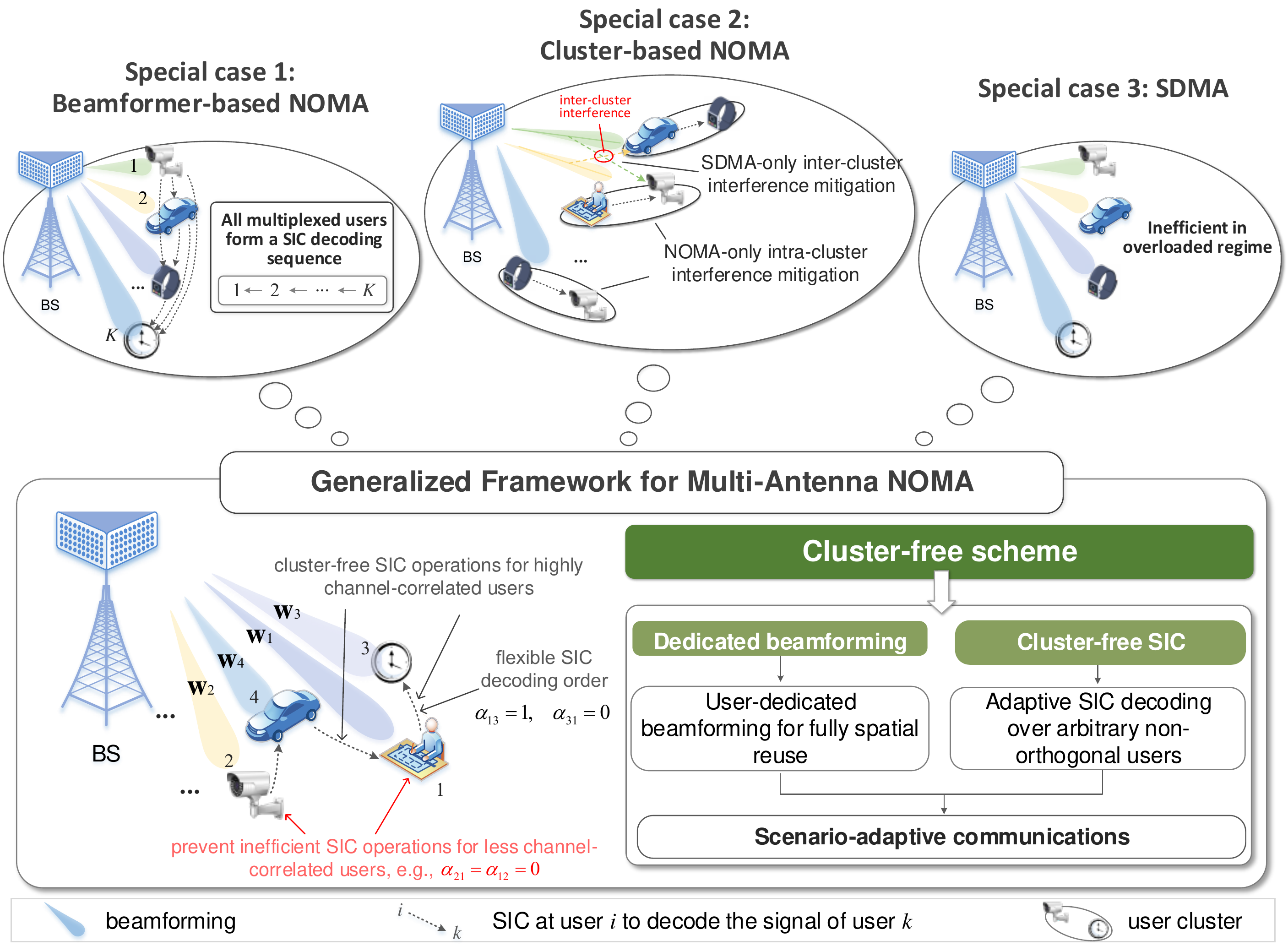}\\
  \caption{Illustration of the proposed generalized cluster-free multi-antenna NOMA framework.}\label{fig_system_model}
  \vspace{-0.3cm}
\end{figure}

We consider a downlink multi-antenna NOMA system, as shown in Fig. \ref{fig_system_model}.
There exists an $M$-antenna base station (BS) serving $K$ single-antenna users randomly distributed within its coverage, indexed by $\mathcal{K}=\{1,2,...,K\}$.
Define $\mathbf{h}_k\in\mathbb{C}^{M\times 1}$ as the channel vector from the BS to user $k$.
The channel gains experience independent and identically distributed (i.i.d) block fading.
Each user $k$ is served by a dedicated transmit beamforming vector $\mathbf{w}_k \in \mathbb{C}^{N\times1}$. Since the proposed framework eliminates the concept of cluster, each user is not required to share beamforming vectors with any other users. 
Denote the transmit beamforming matrix by $\mathbf{W} = \left[\mathbf{w}_1,\mathbf{w}_2,...,\mathbf{w}_K\right]\in\mathbb{C}^{N\times K}$.
For each user $k\in\mathcal{K}$, the received signal can be expressed as
\vspace{-0.5em}
\begin{equation}\label{Signal}
y_{k}\left(\mathbf{W}\right) = \underset{\text{desired signal}}{\underbrace{\mathbf{h}_{k}^{H}\mathbf{w}_{k}s_{k}}}\\
+ \underset{\text{inter-user interference}}{\underbrace{\sum\limits_{ u\ne k}\mathbf{h}_{k}^{H}\mathbf{w}_{u}s_{u}}}+\underset{\text{noise}}{\underbrace{z_{k}}},
\vspace{-0.5em}
\end{equation}
where $s_k$ denotes the data signal of user $k$ with normalized power, i.e., $\mathbb{E}\left\{s_{k}s_{k}^{H}\right\} = 1$. Moreover,
$z_k$ denotes the additive white Gaussian noise (AWGN), which can be modeled as circularly symmetric i.i.d zero-mean complex Gaussian variables, i.e., $z_k \sim \mathcal{CN}\left( 0,\sigma^2 \right)$. %with $\sigma^2$ being the average background noise power.

To efficiently mitigate the inter-user interference suffered by each user, the proposed framework introduces a novel cluster-free SIC concept, which differs from traditional methods in that it enables SIC to be flexibly implemented between any two non-orthogonal users without the predefined user clusters.
Mathematically, we define the binary indicator $\alpha_{ik}$, $\forall i,k \in \mathcal{K}$, which specifies whether the SIC operation is carried out at user $i$ to decode the signal of user $k$.
Specifically, $\alpha_{ik}=1$ indicates that user $i$ will first employ the SIC to decode the signal of user $k$ before decoding its own signal for eliminating interference from user $k$, and $\alpha_{ik}=0$ otherwise.
As it is generally impossible to mutually implement the SIC decoding at both users, we have
\vspace{-0.5em}
\begin{equation}\label{DecodingOrder}
\alpha_{ik} + \alpha_{ki} \le 1, ~ \forall i,k \in \mathcal{K}, ~ i \ne k.
\vspace{-0.5em}
\end{equation}
As implied by \eqref{DecodingOrder}, the variables $\bm{\alpha}$ also determines the SIC decoding order.
The achievable rate of the proposed framework can be modelled as follows.

\subsubsection{Communication rate modelling}
When user $k$ decodes its own signal, the observed interference $\mathrm{Intf}_{k\rightarrow k}$ after SIC operations can be expressed as
\vspace{-0.5em}
\begin{equation}\label{Intf}
\mathrm{Intf}_{k\rightarrow k}\left(\bm{\alpha},\mathbf{W}\right)= \sum\limits_{u\ne k}\left(1-\alpha_{ku}\right)\left|\mathbf{h}_{k}^{H}\mathbf{w}_{u}\right|^{2}+\sigma^{2},
\vspace{-0.5em}
\end{equation}
where $\bm{\alpha}$ is the matrix defined as $\bm{\alpha}=\left[\bm{\alpha}_{1}, ...\bm{\alpha}_{K}\right]$, with $\bm{\alpha}_{k}=\left[\alpha_{1k}, \alpha_{2k},....,\alpha_{Kk}\right]^{T}$ being the SIC operation vector for user $k$.
Therefore, the signal-to-interference-plus-noise ratio (SINR) $\mathrm{SINR}_{i\rightarrow k}$ for user $k$ to decode its own signal can be given by
\vspace{-0.5em}
\begin{equation}\label{SINR}
\mathrm{SINR}_{k\rightarrow k} \left(\bm{\alpha},\mathbf{W}\right)= \frac{\left|\mathbf{h}_{k}^{H}\mathbf{w}_{k}\right|^{2}}{\mathrm{Intf}_{k\rightarrow k}\left(\bm{\alpha},\mathbf{W}\right)}, ~ \forall k\in\mathcal{K}.
\vspace{-0.5em}
\end{equation}
As a result, the achievable data rate $R_{k\rightarrow k}\left(\bm{\alpha},\mathbf{W}\right)$ for user $k$ to decode its own signal can be computed as
$R_{k\rightarrow k}\left(\bm{\alpha},\mathbf{W}\right)= \log_2\left(1+\mathrm{SINR}_{k\rightarrow k} \left(\bm{\alpha},\mathbf{W}\right)\right), ~ \forall k\in\mathcal{K}$.

\begin{figure}[!t]
\vspace{-1.2em}
	\centering
	\subfloat[Normal case.]{\centering \scalebox{0.25}{\includegraphics{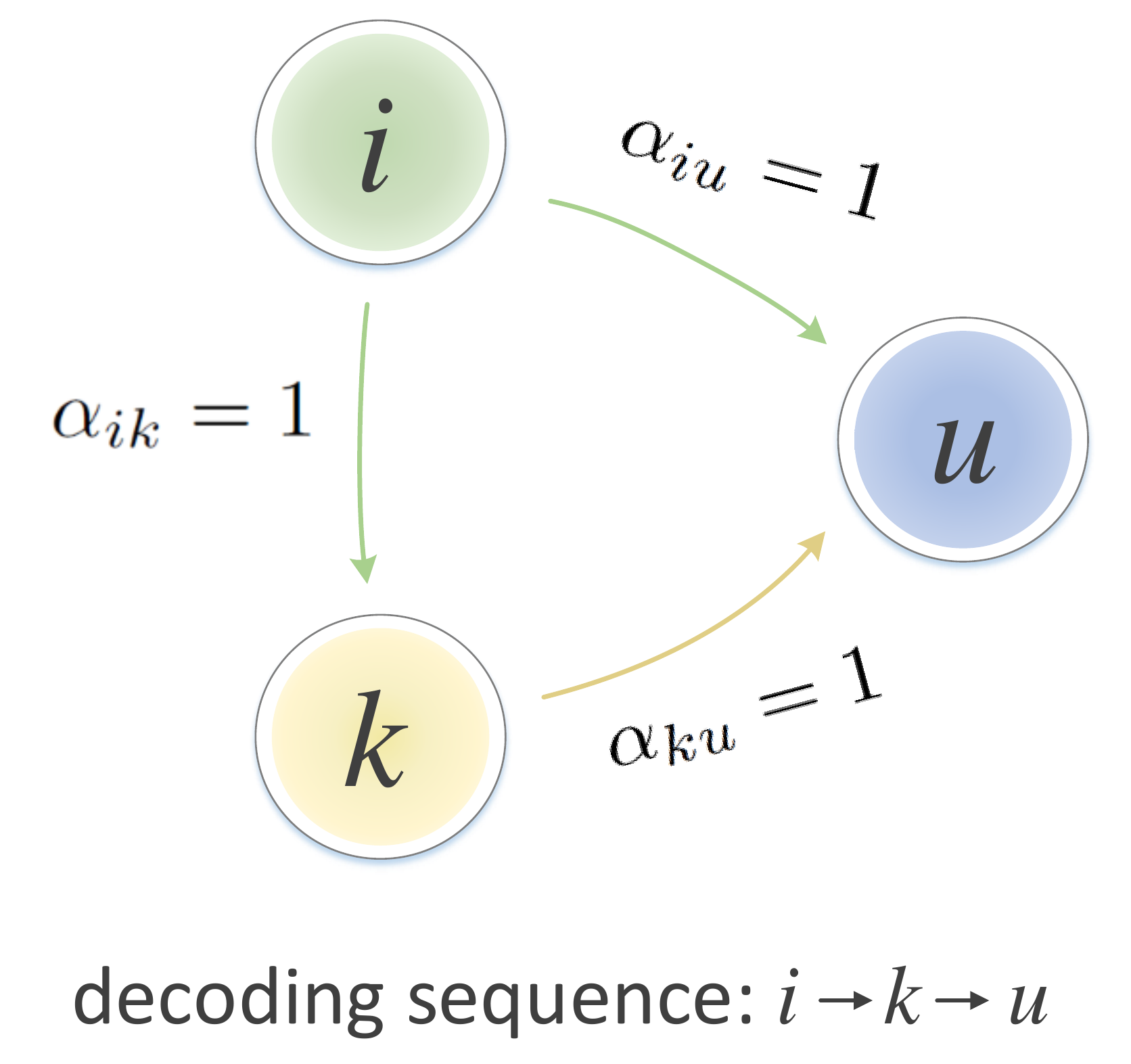}} }
	\quad\quad\quad\quad\quad
    \subfloat[Additional case.]{\centering \scalebox{0.25}{\includegraphics{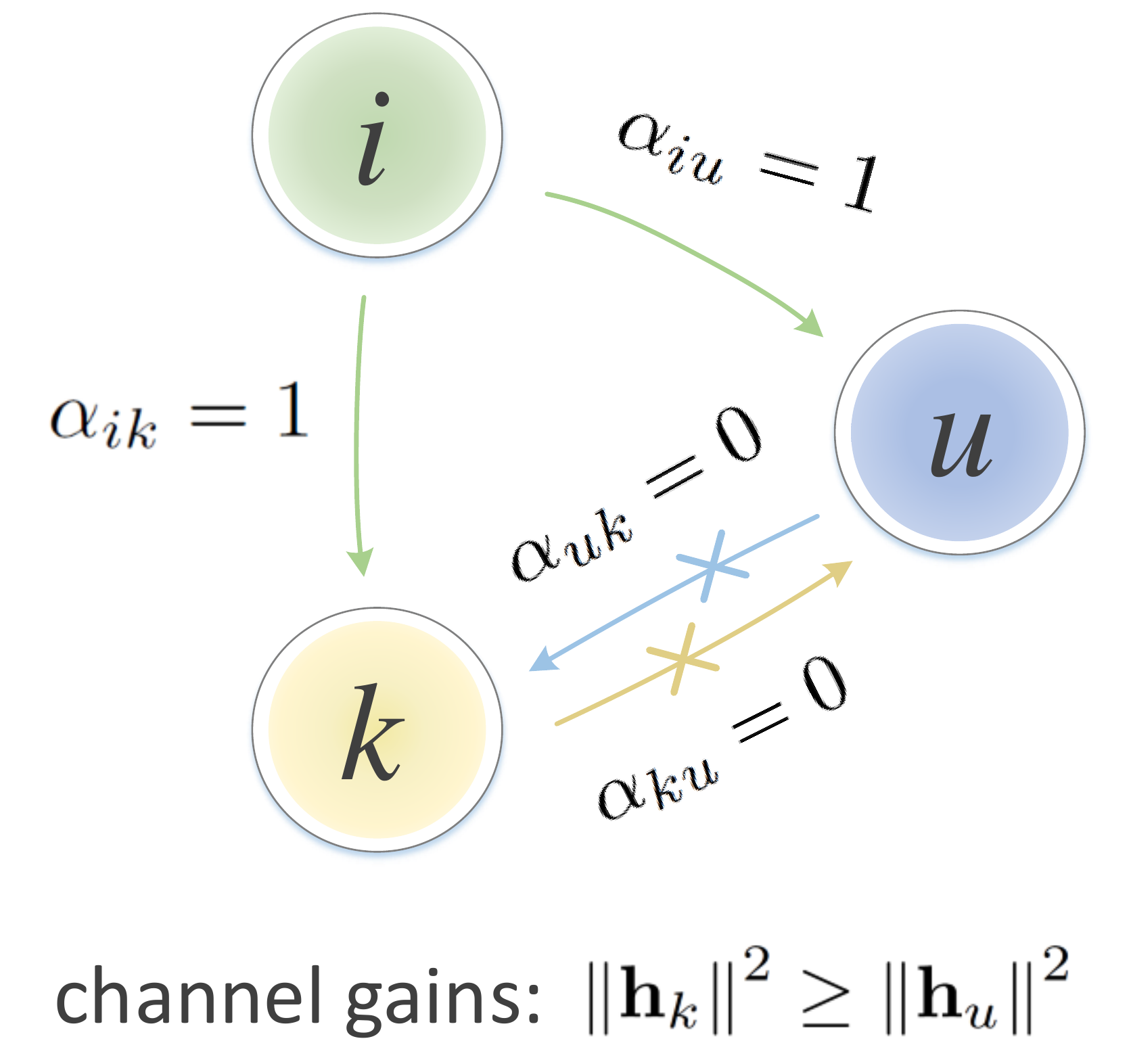}} }	
	\caption{Two cases that interference from user $u$, $u\ne k$, can be eliminated when user $i$ decoding the signal of user $k$ (given that SIC decoding condition \eqref{SICConstraint} is satisfied) : \textbf{(a) the normal case}, which is similar to traditional multi-antenna NOMA, and \textbf{(b) an additional case}, which requires $\left\|\mathbf{h}_{k}\right\|^2\ge\left\|\mathbf{h}_{u}\right\|^2$, $\alpha_{ik}=\alpha_{iu}=1$, and $\alpha_{uk}=\alpha_{ku}=0$.}\label{fig_decoding}
\vspace{-1.2em}
\end{figure}

\subsubsection{SIC decoding rate modelling}
By employing SIC, when $\alpha_{ik}=1$, user $i$ needs to decode user $k$'s signal before decoding its own signal.
Define $\mathrm{Intf}_{i\rightarrow k}\left(\bm{\alpha},\mathbf{W}\right)$, $\forall i,k\in\mathcal{K}$, $i\ne k$, as the observed interference when user $i$ decoding the signal of user $k$.
To model $\mathrm{Intf}_{i\rightarrow k}\left(\bm{\alpha},\mathbf{W}\right)$, we should determine in which cases the interference from the other user $u$, $u\in\mathcal{K}$, $u\ne k$, can be eliminated via SIC when decoding user $k$'s signal at user $i$.
As shown in Fig. \ref{fig_decoding}, this can only happen in the following two cases if all the involved SIC decoding can be successfully carried out:
\begin{itemize}
  \item[(i)] On the one hand, if both user $i$ and user $k$ carry out SIC to decode the signal of user $u$, i.e., $\alpha_{iu}=\alpha_{ku}=1$, the interference $\left|\mathbf{h}_{i}^{H}\mathbf{w}_{u}\right|^2$ from user $u$ can be eliminated when decoding the signal of user $k$ at user $i$.
      As depicted in Fig. \ref{fig_decoding}(a), this case follows the same principle of conventional multi-antenna NOMA, i.e., the three users are regarded in the same cluster.
  \item[(ii)] On the other hand, if no SIC operation is carried out between user $k$ and user $u$, i.e., $\alpha_{ku} = \alpha_{uk} = 0$, and user $i$ employs SIC to decode the signals of both $u$ and $k$, i.e., $\alpha_{iu}=\alpha_{ik}=1$, then user $i$ would sequentially decode signals of user $k$ and user $u$ according to the ascending order of their channel gains. Therefore,when user $i$ decodes user $k$'s signal, the interference from user $u$ can be eliminated via SIC if $\left\|\mathbf{h}_u\right\|^2 \le \left\|\mathbf{h}_k\right\|^2$, as depicted in Fig. \ref{fig_decoding}(b).
\end{itemize}

Without loss of generality, we assume that users are sorted in ascending order of their channel gains, i.e.,  $\left\|\mathbf{h}_u\right\|^2 \le \left\|\mathbf{h}_k\right\|^2$, $\forall u<k$.
Therefore, given $\alpha_{ik}=1$, we can achieve that when $u<k$, for decoding the signal of user $k$ at user $i$, the interference from user $u$ cannot be eliminated if $\alpha_{iu}=0$ or $\alpha_{iu}=\alpha_{uk}=1$. Otherwise, when $u>k$, the interference from user $u$ cannot be eliminated if $\alpha_{iu}=0$ or $\alpha_{ku}=0$. 
This yields the interference $\mathrm{Intf}_{i\rightarrow k}\left(\bm{\alpha},\mathbf{W}\right)$ for decoding the signal of user $k$ at user $i$, which can be mathematically formulated by
\vspace{-0.3em}
\begin{equation}\label{Intf_decoding1}
\mathrm{Intf}_{i\rightarrow k}\left(\bm{\alpha},\mathbf{W}\right) \!=\!
\sum\limits_{u<k}\!\left(1\!-\!\alpha_{iu}\!+\!\alpha_{iu}\alpha_{uk}\right)\!\left|\mathbf{h}_{i}^{H}\mathbf{w}_{u}\right|^{2}\!
+\!\sum\limits_{u>k}\left(1\!-\!\alpha_{iu}\alpha_{ku}\right)\left|\mathbf{h}_{i}^{H}\mathbf{w}_{u}\right|^{2}\!+\!\sigma^{2},
~\forall i,k\in\mathcal{K}, i\ne k.
\vspace{-0.3em}
\end{equation}

The corresponding SINR for user $i$ to decode user $k$'s signal, defined as $\mathrm{SINR}_{i\rightarrow k} \left(\bm{\alpha},\mathbf{W}\right)$, can be computed by
\vspace{-0.5em}
\begin{equation}\label{SINR_decoding1}
\mathrm{SINR}_{i\rightarrow k} \left(\bm{\alpha},\mathbf{W}\right) =
\frac{\left|\mathbf{h}_{i}^{H}\mathbf{w}_{k}\right|^2}{\mathrm{Intf}_{i\rightarrow k}\left(\bm{\alpha},\mathbf{W}\right)}
~\forall i,k\in\mathcal{K}, ~ i\ne k.
\vspace{-0.5em}
\end{equation}
Moreover, the achievable data rate for the SIC decoding can be given by $R_{i\rightarrow k} \left(\bm{\alpha},\mathbf{W}\right) = \left(1+\mathrm{SINR}_{i\rightarrow k} \left(\bm{\alpha},\mathbf{W}\right)\right)$, $\forall i,k\in\mathcal{K}$, $i\ne k$.

To completely eliminate the interference via the SIC as described above, the following condition has to be satisfied to ensure the successful SIC decoding when $\alpha_{ik}=1$ \cite{MIMONOMA_2018}
\vspace{-0.5em}
\begin{equation}\label{SICConstraint}
R_{i\rightarrow k}\left(\bm{\alpha},\mathbf{W}\right) \ge \alpha_{ik} R_{k \rightarrow k}\left(\bm{\alpha},\mathbf{W}\right),
~\forall i,k \in \mathcal{K}, ~i\ne k,
\vspace{-0.5em}
\end{equation}

The sum rate of the proposed generalized cluster-free NOMA framework can be given by
\vspace{-0.5em}
\begin{equation}\label{sumrate}
R\left(\bm{\alpha},\mathbf{W}\right) = \sum\limits_{k\in\mathcal{K}} R_{k \rightarrow k}\left(\bm{\alpha},\mathbf{W}\right).
\vspace{-0.5em}
\end{equation}

Essentially, by introducing the cluster-free SIC, the proposed framework provides a generalized and unified modelling, where the beamformer-based NOMA, cluster-based NOMA, and SDMA can be all regarded as the special cases of the proposed framework, as analysed as follows.

\textit{1) Special case 1 - Beamformer-based NOMA:}
When there is only one SIC decoding sequence that involves all the connected users, i.e., $\alpha_{ik}=1$, $\forall i>k$, and $\alpha_{ik}=0$ otherwise,
the proposed generalized framework is equivalent to beamformer-based NOMA.
In this case, the achievable sum rate can be given by
%\vspace{-0.5em}
$
R^{\mathrm{BB-NOMA}}=\sum\limits_{k\in\mathcal{K}}\log_2\left(1+\frac{\left|\mathbf{h}_{k}^{H}\mathbf{w}_{k}\right|^{2}}{\sum\limits_{k< u}\left|\mathbf{h}_{k}^{H}\mathbf{w}_{u}\right|^{2}+\sigma^{2}}\right).
%\vspace{-0.5em}
$

\textit{2) Special case 2 - Cluster-based NOMA:}
When user $i$ and user $k$ are served by two aligned beamforming vectors, i.e., $\exists c_{ik}\in\mathbb{R}$, $\mathbf{w}_k = c_{ik} \mathbf{w}_j$, $i,k \in \mathcal{K}$,  they are considered to share the same spatial beam, which is similar to the traditional cluster-based NOMA systems. 
If SIC decoding is sequentially carried out only between users served by aligned beamforming vectors, then the proposed generalized framework reduces to cluster-based NOMA, where
$\alpha_{ik}=1$ if $i>k$ and $\exists c_{ik}\in\mathbb{R}$ such that $\mathbf{w}_{k}=c_{ik}\mathbf{w}_{i}$, and $\alpha_{ik}=0$ otherwise.
Suppose there exists $G$ clusters, indexed by $\mathcal{G}=\{1,2,...,G\}$.
Denote the user set of cluster $g$ by $\mathcal{K}_{g}$.
Then, the achievable sum rate of the cluster-based NOMA system can be written as
%\vspace{-0.3em}
$
R^{\mathrm{CB-NOMA}} =
\sum\limits_{k\in\mathcal{K}}\log_2\left(1+\frac{\left|\mathbf{h}_{k}^{H}\mathbf{w}_{k}\right|^{2}}
{\sum\limits_{k\in\mathcal{K}_g,k<u}\left|\mathbf{h}_{k}^{H}\mathbf{w}_{u}\right|^{2}
+\sum\limits_{u\notin\mathcal{K}_g}\left|\mathbf{h}_{k}^{H}\mathbf{w}_{u}\right|^{2}+\sigma^{2}} \right).
\vspace{0.2em}
$

\textit{3) Special case 3 - SDMA:}
If there is no SIC operation between any users, i.e., $\alpha_{ik}= 0, ~ \forall i,k \in\mathcal{K}$, $i\ne k,$ then the proposed generalised NOMA framework is equivalent to SDMA.
The sum rate of the SDMA system can be expressed as
$
R^{\mathrm{SDMA}}=\sum\limits_{k\in\mathcal{K}}\log_2\left(1+\frac{\left|\mathbf{h}_{k}^{H}\mathbf{w}_{k}\right|^{2}}{\sum\limits_{k\ne  u}\left|\mathbf{h}_{k}^{H}\mathbf{w}_{u}\right|^{2}+\sigma^{2}}\right).
$

In addition to unifying the traditional methods, the proposed framework also enables more flexible SIC operations.
A specific example is shown in Fig. \ref{fig_system_model}, where users cannot be ideally divided into a single or multiple user clusters, and the cluster-specific SIC schemes are not flexible enough. To empower efficient interference elimination, the proposed cluster-free scheme breaks the clustering limitations, which can flexibly enables SIC operations between highly channel-correlated users (e.g., user $2$ and user $4$, user $1$ and user $4$), while adaptively preventing ineffective SIC operations between less channel-correlated users (e.g., user $1$ and user $2$).

\vspace{-0.5em}
\begin{remark}\label{remark_generalization}
The proposed framework provides a generalized model to unify traditional methods, and enables more flexible transmission options with cluster-free SIC to achieve adaptive inter-user interference mitigation.
Therefore, it can overcome the defects of traditional methods and reap their gains to deal with diverse scenarios facing next-generation wireless communications.
Owing to these merits, we can straightforwardly derive that the achievable sum rate of the proposed framework can outperform or is at least not worse than traditional approaches, i.e.,
\vspace{-0.5em}
\begin{equation}\label{rate_comparison}
R\left(\bm{\alpha},\mathbf{W}\right) \ge \max\{R^{\mathrm{BB-NOMA}}, R^{\mathrm{CB-NOMA}},R^{\mathrm{SDMA}}\}.
\vspace{-0.6em}
\end{equation}
\end{remark}

\vspace{-0.8em}
\subsection{Problem Formulation}
Our goal is to maximize the sum rate while guaranteeing SIC decoding conditions and ensuring users' data rate requirements by jointly optimizing the transmit beamforming and the cluster-free SIC operations between users.
Mathematically, the optimization problem can be formulated as\footnote{Considering the fact that each user would always decode its own signal, we directly set $\mathrm{diag}\left(\bm{\alpha}\right)= \left\{\alpha_{i,i}\right\} = \mathbf{1}_{K \times 1}$ and optimize variables $\alpha_{ik}$, ${\forall i \ne k}$, here. }
\vspace{-0.3em}
\begin{subequations}\label{P0}
\begin{align*}
\mathcal{P}_{0}: & \max_{\bm{\alpha},\mathbf{W}} ~
\sum\limits_{k\in\mathcal{K}}\log_2 \left(1+\mathrm{SINR}_{k\rightarrow k}\left(\bm{\alpha},\mathbf{W}\right)\right)  \label{P0_obj} \tag{\ref{P0}{a}}
\\ {\mathrm{s.t.}}~ &
\log_2\left(1+\mathrm{SINR}_{i\rightarrow k}\left(\bm{\alpha},\mathbf{W}\right)\right)
\ge \alpha_{ik}\log_2\left(1+\mathrm{SINR}_{k\rightarrow k}\left(\bm{\alpha},\mathbf{W}\right)\right), ~ \forall i,k \in \mathcal{K}, \label{constraint_SIC} \tag{\ref{P0}{b}}
\\&
\log_2\left(1+\mathrm{SINR}_{k\rightarrow k}\left(\bm{\alpha},\mathbf{W}\right)\right)
\ge R_{k}^{\min}, ~ \forall k \in \mathcal{K}, \label{constraint_rate} \tag{\ref{P0}{c}}
\\&
\sum\limits_{k\in\mathcal{K}} \left\|\mathbf{w}_k\right\|^2 \le P^{\max}, \label{constraint_power} \tag{\ref{P0}d}
\\&
\alpha_{ik} + \alpha_{ki} \le 1, ~\forall i,k \in \mathcal{K}, ~ i\ne k,  \label{constraint_Decoding} \tag{\ref{P0}e}
\\&
\alpha_{ik} \in\{0, 1\}, ~ \forall i,k \in \mathcal{K}, \label{constraint_bin} \tag{\ref{P0}f}
\end{align*}
\end{subequations}
\vspace{-1cm}

\noindent where constraint \eqref{constraint_SIC} represents the SIC decoding conditions rearranged from \eqref{SICConstraint},
\eqref{constraint_rate} guarantees the minimum data rate of each user $i$,
and \eqref{constraint_power} ensures the maximum transmit power of BS does not exceed $P^{\max}$.
Furthermore, \eqref{constraint_Decoding} indicates that user $i$ and user $k$, $i\ne k$, cannot mutually implement the SIC decoding,
and \eqref{constraint_bin} indicates the binary variable constraint.

Nevertheless, it's challenging to solve $\mathcal{P}_{0}$ owing to the following reasons.
Firstly, the SINR expressions in \eqref{P0_obj}-\eqref{constraint_rate} are neither convex nor concave with respect to the optimization variables.
Additionally, the design of SIC operations introduces the binary constraint \eqref{constraint_bin}.
Furthermore, the optimization variables are highly coupled with each others in both the interference terms and the objective function.
Therefore, $\mathcal{P}_{0}$ is a non-convex and highly coupled MINLP problem, which is nondeterministic polynomial time-hard (NP)-hard. This makes it difficult to find the globally optimal solution.
To deal with these difficulties, locally algorithms are proposed in the following sections.

\section{ADMM-SCA Based Solution}
In this section, an ADMM-SCA algorithm is developed to solve $\mathcal{P}_{0}$.
The highly coupled MINLP is first equivalently reformulated into a tractable AL problem with continuous variables.
By invoking the SCA to handle the non-convex terms, the AL problem can be approximately transformed into a series of biconvex optimization problem.
Based on the strongly convergence-guaranteed ADMM method, we further decompose the biconvex problem into two convex subproblems, which can be iteratively solved to achieve the stationary solution.

\vspace{-0.8em}
\subsection{Problem Transformation}
To tackle the original NP-hard MINLP problem \eqref{P0}, we first introduce the auxiliary variables $\bm{\beta}\in\mathbb{R}^{K \times K}$, which satisfy
\vspace{-0.8em}
\begin{equation}\label{constraint_beta2}
\beta_{ik} = 1 - \alpha_{ik}, ~ \forall i, k, \in \mathcal{K},
\vspace{-0.8em}
\end{equation}
\begin{equation}\label{constraint_bin2}
\alpha_{ik} \beta_{ik} = 0, ~ \forall i, k, \in \mathcal{K},
\vspace{-0.8em}
\end{equation}
\begin{equation}\label{constraint_var2}
0 \le \alpha_{ik}, \beta_{ik} \le 1, ~ \forall i, k, \in \mathcal{K}.
\vspace{-0.8em}
\end{equation}
Since constraints \eqref{constraint_beta2}-\eqref{constraint_var2} enforces $\alpha_{ik}\left(1-\alpha_{ik}\right) = 0$ and $\beta_{ik}\left(1-\beta_{ik}\right) = 0$, they can stringently guarantee that $\alpha_{ik},\beta_{ik} \in\{0,1\}$.
Therefore, the discrete binary constraint \eqref{constraint_bin} can be equivalently replaced by \eqref{constraint_beta2}-\eqref{constraint_var2}.

Moreover, since the interference term $\mathrm{Intf}_{i\rightarrow k}\left(\bm{\alpha},\mathbf{W}\right)$ in \eqref{Intf_decoding1} suffers from the highly  coupling variables $\alpha_{ki}$, $\alpha_{iu}$, $\alpha_{uk}$, and $\mathbf{W}$, we equivalently transform the interference term as follows to make it tractable.
Since both $\left\{\alpha_{ik}\right\}$ and $\left\{\beta_{ik}\right\}$ are binary variables, the coupling terms $\left(1-\alpha_{iu}+\alpha_{iu}\alpha_{uk}\right)$ and $\left(1-\alpha_{iu}\alpha_{ku}\right)$, $\forall i,u,k\in\mathcal{K}$, in \eqref{Intf_decoding1} can be directly  recast as
% \vspace{-0.5em}
\begin{equation}\label{alp_bet_1}
1-\alpha_{iu}+\alpha_{iu}\alpha_{uk}
=\max\left\{ 1\!-\!\alpha_{iu},\alpha_{uk}\right\}
=\max\left\{ \beta_{iu},1\!-\!\beta_{uk}\right\} ,
\vspace{-0.5em}
\end{equation}
\begin{equation}\label{alp_bet_2}
\left(1-\alpha_{iu}\alpha_{ku}\right)
=\max\left\{1-\alpha_{iu},1-\alpha_{ku}\right\}
=\max\left\{\beta_{iu},\beta_{ku}\right\}.
\vspace{-0.5em}
\end{equation}
Therefore, the interference terms $\mathrm{Intf}_{k\rightarrow k}\left(\bm{\alpha},\mathbf{W}\right)$, $\forall k \in\mathcal{K}$, in \eqref{Intf} and $\mathrm{Intf}_{i\rightarrow k}\left(\bm{\alpha},\mathbf{W}\right)$, $\forall i,k \in\mathcal{K}$, $i\ne k$, in \eqref{Intf_decoding1} can be equivalently rewritten as
\vspace{-0.5em}
\begin{equation}\label{Intf_beta}
\begin{split}
&\widetilde{\mathrm{Intf}}_{i\rightarrow k}\left(\bm{\beta},\mathbf{W}\right) \!=\!
\begin{cases}
\sum\limits_{u\ne k}\!\beta_{ku}\left|\mathbf{h}_{k}^{H}\mathbf{w}_{u}\right|^{2}+\sigma^{2}, \!&\!\! i=k,\\
\sum\limits_{u<k}\!\max\left\{ \beta_{iu},\!1\!-\!\beta_{uk}\right\} \left|\mathbf{h}_{i}^{H}\mathbf{w}_{u}\right|^{2}\!+\!\sum\limits_{u>k}\!\max\!\left\{ \beta_{iu},\!\beta_{ku}\right\} \!\left|\mathbf{h}_{i}^{H}\mathbf{w}_{u}\right|^{2}\!+\!\sigma^{2}, \!&\!\! i\ne k.
\end{cases}
\end{split}
\vspace{-1.2em}
\end{equation}
\vspace{-0.4cm}

\vspace{-0.8cm}
\begin{lemma}\label{remark:convexity_intf}
The function $\widetilde{\mathrm{Intf}}_{i\rightarrow k}\left(\bm{\beta},\mathbf{W}\right)$ defined by \eqref{Intf_beta} is convex with respect to $\bm{\beta}$.
\begin{proof}
For $i=k$, $\widetilde{\mathrm{Intf}}_{i\rightarrow k}\left(\bm{\beta},\mathbf{W}\right)$ is a linear function of $\bm{\beta}$.
Therefore, we only need to verify the convexity for $i\ne k$.
According to the derivation in \cite{ConvexOpt}, the pointwise maximum function $g(\bm{\beta}) = \max\{g_1(\bm{\beta}),g_2(\bm{\beta})\}$ is convex if $g_1(\bm{\beta})$ and $g_2(\bm{\beta})$ are both convex functions.
Since \eqref{alp_bet_1} and \eqref{alp_bet_2} are both pointwise maximums of affine functions of $\bm{\beta}$, it can be concluded that $\widetilde{\mathrm{Intf}}_{i\rightarrow k}$ is convex with respect to $\bm{\beta}$, which completes the proof.
\end{proof}
\vspace{-0.2cm}
\end{lemma}

To deal with the non-convex data rate expression, we further introduce a series of auxiliary variables $\mathbf{S}=\left\{S_{ik}\right\}_{\forall i,k\in\mathcal{K}}$, $\mathbf{I}=\left\{I_{ik}\right\}_{\forall i,k\in\mathcal{K}}$, and $\mathbf{r}=\left\{r_{ik}\right\}_{\forall i,k\in\mathcal{K}}$.
Specifically, $I_{ik}$ indicates the upper bound of the interference $\widetilde{\mathrm{Intf}}_{i\rightarrow k}\left(\bm{\beta},\mathbf{W}\right)$, $\forall i,k \in\mathcal{K}$.
Moreover, $S_{ik}$ and $r_{ik}$ signify the lower bounds of the effective gains and the achievable rate for decoding user $k$'s signal at user $i$, $\forall i,k \in\mathcal{K}$, respectively.
Therefore, the intractable MINLP problem \eqref{P0} can be written as the following continuous problem:
\vspace{-0.7em}
\begin{subequations}\label{P2}
\begin{align*}
\mathcal{P}_{1}: & \max_{\bm{\alpha},\bm{\beta},\mathbf{W},\mathbf{S},\mathbf{I},\mathbf{r}} ~ \sum\limits_{k\in\mathcal{K}}r_{kk} \tag{\ref{P2}{a}}
\\ {\mathrm{s.t.}}~ &
r_{ik}\le\log_{2}\left(1+\frac{S_{ik}}{I_{ik}}\right), ~ \forall i,k\in\mathcal{K}, \label{constraint_rub2} \tag{\ref{P2}{b}}
\\&
S_{ik}\le \left|\mathbf{h}_{i}^H\mathbf{w}_{k}\right|^2, ~ \forall i,k\in\mathcal{K}, \label{constraint_gain2} \tag{\ref{P2}{c}}
\\&
\widetilde{\mathrm{Intf}}_{i\rightarrow k}\left(\bm{\beta},\mathbf{W}\right) \le  I_{ik},
~ \forall i,k\in\mathcal{K}, \label{constraint_intf2} \tag{\ref{P2}{d}}
\\&
r_{ik}\ge \alpha_{ik}r_{kk}, ~\forall i,k \in\mathcal{K}, ~ i \ne k,  \label{constraint_rlb2} \tag{\ref{P2}{e}}
\\&
r_{kk} \ge r_{k}^{\min}, ~\forall k \in\mathcal{K}, \label{constraint_rate2} \tag{\ref{P2}{f}}
\\&
\alpha_{ik} + \alpha_{ki} \le 1, ~\forall i,k \in \mathcal{K}, ~ i\ne k, \label{constraint_alpha2} \tag{\ref{P2}g}
\\&
\sum\limits_{k\in\mathcal{K}}\left\|\mathbf{w}_k\right\|^2 \le P^{\max}, \label{constraint_power2}  \tag{\ref{P2}h}
\\&
\eqref{constraint_beta2}-\eqref{constraint_var2}.  \tag{\ref{P2}i}
\end{align*}
\vspace{-1cm}
\end{subequations}

\begin{proposition}
Problems $\mathcal{P}_{1}$ and $\mathcal{P}_{0}$ are equivalent in the sense that they have equivalent optimal solutions.
\begin{proof}
Owing to the monotonicity of the $\log_2(\cdot)$ function, the constraints \eqref{constraint_rub2}-\eqref{constraint_intf2} in problem $\mathcal{P}_{1}$ always hold with equality at the optimum point.
Therefore, the solutions obtained by solving problem $\mathcal{P}_{1}$ can satisfy $r_{ik}^{*}=\log_2 \left(1+\frac{\left|\mathbf{h}_{k}^{H}\mathbf{w}_{k}^{*}\right|^{2}}{\widetilde{\mathrm{Intf}}_{i\rightarrow k}\left(\bm{\beta},\mathbf{W}^{*}\right)}\right)=\log_2 \left(1+\mathrm{SINR}_{i\rightarrow k}^{*}\left(\bm{\alpha}^{*},\mathbf{W}^{*}\right)\right)$, $\forall i,k \in \mathcal{K}$, which demonstrates the equivalence between the optimal solutions of $\mathcal{P}_{1}$ and  $\mathcal{P}_{0}$.
\end{proof}
%\vspace{-0.6em}
\end{proposition}
\vspace{-0.2cm}

Now we can invoke the strongly convergence-guaranteed ADMM framework \cite{ADMM_1976} to deal with the resulting problem $\mathcal{P}_{1}$.
By dualizing and penalizing the coupling equality constraints \eqref{constraint_beta2} and \eqref{constraint_bin2} into the objective function, the AL problem of $\mathcal{P}_1$ can be formulated as \cite{ADMM_Boyd}

\vspace{-1.5em}
\begin{subequations}\label{PAL}
\begin{align*}
\mathcal{P}_{\mathrm{AL}}:& \max_{\bm{\alpha},\bm{\beta},\mathbf{W},\mathbf{S},\mathbf{I},\mathbf{r}}
~ f_0\left(\mathbf{r}\right)
-\mathcal{L}^{(1)}\left(\bm{\alpha},\bm{\beta},\bm{\lambda}\right) -\mathcal{L}^{(2)}\left(\bm{\alpha},\bm{\beta},\bm{\widetilde{\lambda}}\right) \label{PAL_obj}\tag{\ref{PAL}{a}}
\\ {\mathrm{s.t.}}~ &
\eqref{constraint_rub2} - \eqref{constraint_power2}, \eqref{constraint_var2},  \label{constraint_varAL} \tag{\ref{PAL}{b}}
\end{align*}
\end{subequations}
\vspace{-0.8cm}

\noindent where $f_0\left(\mathbf{r}\right)\!=\!\sum\limits_{k\in\mathcal{K}} \!\log_2\left(1+r_{kk}\right)$ is the original objective function.
$\mathcal{L}^{(1)}\left(\bm{\alpha},\bm{\beta},\bm{\lambda}\right)$ and $\mathcal{L}^{(2)}\!\left(\!\bm{\alpha},\bm{\beta},\bm{\widetilde{\lambda}}\right)\!$ respectively denote the AL terms corresponding to equality constraints \eqref{constraint_beta2} and \eqref{constraint_bin2}, given by
% \vspace{-0.5em}
\begin{equation}\label{AL1}
\mathcal{L}^{(1)}\left(\bm{\alpha},\bm{\beta},\bm{\lambda}\right)=\frac{1}{2\rho}\left\Vert \bm{\beta}+\mathbf{\bm{\alpha}}-\mathbf{1}_{K\times K}+\rho\bm{\lambda}\right\Vert^{2},
\vspace{-0.5em}
\end{equation}
\begin{equation}\label{AL2}
\mathcal{L}^{(2)}\left(\bm{\alpha},\bm{\beta},\bm{\widetilde{\lambda}}\right)=\frac{1}{2\rho}\sum\limits_{k\in\mathcal{K}}\sum\limits_{i\in\mathcal{K}}\left(\alpha_{ik}\beta_{ik}+\rho\widetilde{\lambda}_{ik}\right)^{2},
% \vspace{-0.5em}
\end{equation}
where $\bm{\lambda}=\big\{\lambda_{ik}\big\}$ and $\bm{\widetilde{\lambda}} = \big\{\widetilde{\lambda}_{ik}\big\}$ are the dual variables and $\rho$ is the non-negative penalty parameter.
As proven in \cite{ADMM_Boyd}, by alternatively optimizing the primal variables $\left\{\bm{\alpha},\bm{\beta},\mathbf{W},\mathbf{S},\mathbf{I},\mathbf{r}\right\}$ and dual variables $\left\{\bm{\lambda},\bm{\widetilde{\lambda}}\right\}$ of the AL problem,
the residuals of constraints \eqref{constraint_beta2} and \eqref{constraint_bin2}, i.e., $\left(\beta_{ik}+\alpha_{ik}-1\right)$ and $\alpha_{ik}\beta_{ik}$, $\forall i,k$, will converge to zeros and the binary constraint can be satisfied.

\vspace{-0.5em}
\subsection{ADMM-SCA Algorithm}\label{Section_ADMM_SCA}
According to \textbf{Lemma \ref{remark:convexity_intf}}, the AL problem $\mathcal{P}_{\mathrm{AL}}$ is convex over $\bm{\beta}$.
However, constraint \eqref{constraint_rub2} is non-convex since $\log_2\left(1+\frac{S_{ik}}{I_{ik}}\right)=\log_2\left(S_{ik}+I_{ik}\right)-\log_2\left(I_{ik}\right)$ is a difference of concave function over $I_{ik}$.
Furthermore, constraint \eqref{constraint_gain2} is non-convex over $\mathbf{W}$.
To handel these non-convex constraints, we integrate the SCA method \cite{SCA_1978} into the ADMM framework \cite{ADMM_1976}.
Utilizing SCA, the non-convex components can be approximately and sequentially linearized  into a series of convex expressions based on the first-order Taylor approximation for a given local point.
Thus, the AL problem can be further decomposed into convex subproblems that can be optimized based on ADMM in an alternative and iterative manner.

Let $\overline{\mathbf{w}}_{k}, \overline{I}_{ik}, \overline{\alpha}_{ik}$, and $\overline{r}_{kk}$ denote the values of the optimization variables $\mathbf{w}_{k}$, $I_{ik}$, $\alpha_{ik}$, and $r_{kk}$ obtained from the previous SCA iteration, respectively.
We first define function $q_{1}\left(\mathbf{w}_{k}\right)=\left|\mathbf{h}_{i}^{H}\mathbf{w}_{k}\right|^{2}$.
Based on the first-order Taylor approximation around $\overline{\mathbf{w}}_k$, i.e., $q_1\left(\mathbf{w}_{k}\right)\ge  \widehat{q}_{1}\left(\mathbf{w}_{k},\overline{\mathbf{w}}_{k}\right) = q_{1}\left(\overline{\mathbf{w}}_{k}\right)
+q_{1}'\left(\mathbf{\overline{w}}_{k}\right)\left(\mathbf{w}_{k}-\mathbf{\overline{w}}_{k}\right)$,
we can recast the constraint \eqref{constraint_gain2} as
\vspace{-0.6em}
\begin{equation}\label{constraint_signalSCA}
S_{ik} +\left|\mathbf{h}_{i}^{H}\mathbf{\overline{w}}_{k}\right|^{2}
\le 2\mathscr{R}\left(\overline{\mathbf{w}}_{k}^{H}\mathbf{h}_{i}\mathbf{h}_{i}^{H}\mathbf{w}_{k}\right),
~ \forall i,k\in\mathcal{K},
\vspace{-0.4em}
\end{equation}

Similarly, by taking the first-order Taylor expansion of function $q_{2}\left(I_{ik}\right)=\log_{2}\left(I_{ik}\right)$ at point $\overline{I}_{ik}$, we can obtain
\vspace{-1em}
\begin{equation}\label{Tylor_I}
q_{2}\left(I_{ik}\right) \le \widehat{q}_{2}\left(I_{ik},\overline{I}_{ik}\right) = \log_{2}\left(\overline{I}_{ik}\right)+\frac{1}{\ln2}\frac{1}{\overline{I}_{ik}}\left(I_{ik}-\overline{I}_{ik}\right).
\vspace{-0.5em}
\end{equation}
After rearrangement, the constraint \eqref{constraint_rub2} can be transferred into
\vspace{-0.4em}
\begin{equation}\label{constraint_rubSCA}
r_{ik}+\log_{2}\left(\overline{I}_{ik}\right)+\frac{1}{\ln2}\frac{1}{\overline{I}_{ik}}\left(I_{ik}-\overline{I}_{ik}\right)
\le \log_{2}\left(I_{ik}+S_{ik}\right), ~ \forall i,k\in\mathcal{K}.
\vspace{-0.4em}
\end{equation}

Furthermore, to decouple $\alpha_{ik}$ and $r_{kk}$ in constraint \eqref{constraint_rlb2}, the term $\alpha_{ik}r_{kk}$ can be rearranged as $\alpha_{ik}r_{kk} = \frac{1}{4}\left(\alpha_{ik}+r_{kk}\right)^{2}-\frac{1}{4}\left(\alpha_{ik}-r_{kk}\right)^{2}$.
To deal with this difference-of-convex expression, we linearize the non-convex term $-\frac{1}{4}\left(\alpha_{ik}-r_{kk}\right)^{2}$ using the first-order Taylor expansion, i.e.,
\vspace{-0.4em}
\begin{equation}\label{Tylor_r}
-\frac{1}{4}\left(\alpha_{ik}-r_{kk}\right)^{2}
\le -\frac{1}{4}\left(\overline{\alpha}_{ik}-\overline{r}_{kk}\right)^{2}
-\frac{1}{2}\left(\overline{\alpha}_{ik}-\overline{r}_{kk}\right)
\times \left(\alpha_{ik}-\overline{\alpha}_{ik}+\overline{r}_{kk}-r_{kk}\right).
\vspace{-0.4em}
\end{equation}
Considering \eqref{Tylor_r}, constraint \eqref{constraint_rlb2} can be transformed into
\vspace{-0.9em}
\begin{equation}\label{constraint_rlbSCA}
r_{ik}\!+\!\frac{1}{4}\left(\overline{\alpha}_{ik}\!-\!\overline{r}_{kk}\right)^{2}\!+\!\frac{1}{2}\left(\overline{\alpha}_{ik}-\overline{r}_{kk}\right)
\left(\alpha_{ik}\!-\!\overline{\alpha}_{ik}\!+\!\overline{r}_{kk}\!-\!r_{kk}\right)
\!\ge\! \frac{1}{4}\left(\alpha_{ik}\!+\!r_{kk}\right)^{2},
~ \forall i,k\in\mathcal{K}, ~ i\ne k,
\vspace{-0.9em}
\end{equation}

Based on the above analyses, the AL problem $\mathcal{P}_{\mathrm{AL}}$ can be approximately transformed into the following problem during each SCA update:
\vspace{-0.7em}
\begin{subequations}\label{PSCA}
\begin{align*}
\mathcal{P}_{2}:& \max_{\bm{\alpha},\bm{\beta},\mathbf{W},\mathbf{I},\mathbf{S,\mathbf{r}}} ~
f_0\left( \mathbf{r} \right)
- \mathcal{L}^{(1)}\left(\bm{\alpha},\bm{\beta},\bm{\lambda}\right) - \mathcal{L}^{(2)}\left(\bm{\alpha},\bm{\beta},\bm{\widetilde{\lambda}}\right) \tag{\ref{PSCA}{a}}
\\ {\mathrm{s.t.}}~ &
\eqref{constraint_var2},  \eqref{constraint_intf2}, \eqref{constraint_rate2} - \eqref{constraint_power2}, \eqref{constraint_signalSCA}, \eqref{constraint_rubSCA}, \eqref{constraint_rlbSCA}.  \tag{\ref{PSCA}{b}}
\vspace{-0.8cm}
\end{align*}
\end{subequations}
\vspace{-1cm}

The resulting problem $\mathcal{P}_{2}$ is a biconvex problem, which can be decomposed into two nested convex subproblems over two variable blocks $\left\{\left\{\alpha_{ik}\right\}_{\forall i \ne k},\mathbf{W}\right\}$ and $\left\{\beta_{ik}\right\}_{\forall i \ne k}$.
Based on the ADMM framework, these convex subproblems can be solved alternatively at each iteration, followed by which the dual variables $\bm{\lambda}$ and $\bm{\widetilde{\lambda}}$ are updated.
In light of this, we propose an ADMM-SCA algorithm, which has three steps during each iteration.

Firstly, given $\left\{\bm{\beta},\overline{\bm{\alpha}},\overline{\mathbf{W}},\overline{\mathbf{I}},\overline{\mathbf{r}},\bm{\lambda},\widetilde{\bm{\lambda}}\right\}$, the ADMM-SCA algorithm jointly optimizes the SIC operations $\left\{\alpha_{ik}\right\}_{\forall i \ne k}$, and the transmit beamforming $\mathbf{W}$ by solving the following convex problem
\vspace{-2em}
\begin{subequations}\label{Palpha}
\begin{align*}
& \max_{\bm{\alpha},\mathbf{W},\mathbf{S},\mathbf{I},\mathbf{r}} ~
f_0\left(\mathbf{r}\right)
- \mathcal{L}^{(1)}\left(\bm{\alpha},\bm{\beta},\bm{\lambda}\right) - \mathcal{L}^{(2)}\left(\bm{\alpha},\bm{\beta},\bm{\widetilde{\lambda}}\right) \tag{\ref{Palpha}{a}}
\\{\mathrm{s.t.}}~ &
\eqref{constraint_var2}, \eqref{constraint_intf2}, \eqref{constraint_rate2} - \eqref{constraint_power2}, \eqref{constraint_signalSCA}, \eqref{constraint_rubSCA}, \eqref{constraint_rlbSCA}.  \label{constraint_varAL} \tag{\ref{Palpha}{b}}
\vspace{-1cm}
\end{align*}
\end{subequations}

\noindent Based on the first-order optimality, the Karush-Kuhn-Tucker (KKT) solution of $\left\{\alpha_{ik}\right\}_{\forall i \ne k}$ is given by
\vspace{-0.6em}
\begin{multline}\label{Sol_alpha}
\alpha_{ik}^{*} =
\frac{\rho\omega_{ik}^{\mathrm{(4)}}\left(\overline{\alpha}_{ik}-\overline{r}_{kk}-r_{kk}\right)
+2\left(1-\beta_{ik}-\rho\lambda_{ik}-\rho\widetilde{\lambda}_{ik}\right)+2\rho\left(\omega_{ik}^{\mathrm{(2)}}-\omega_{ik}^{\mathrm{(1)}}-\omega_{ik}^{(3)}\right)}
{2\left(1+\beta_{ik}\right)+\rho\omega_{ik}^{(4)}},
\vspace{-0.6em}
\end{multline}
where $\omega_{ik}^{\mathrm{(1)}}$ and $\omega_{ik}^{\mathrm{(2)}}$ denotes the Lagrangian multipliers for constraints \eqref{constraint_var2}, and $\omega_{ik}^{(3)}$ and $\omega_{ik}^{(4)}$ are the Lagrangian multipliers corresponding to constraints \eqref{constraint_alpha2} and \eqref{constraint_rlbSCA}, respectively.

Thereafter, we update $\bm{\beta}$ by solving the following problem with fixed $\big\{\bm{\alpha},\mathbf{W},\mathbf{S},\overline{\bm{\alpha}},\overline{\mathbf{I}},\overline{\mathbf{r}},\bm{\lambda},\widetilde{\bm{\lambda}}\big\}$

\vspace{-2.2em}
\begin{subequations}\label{Pbeta}
\begin{align*}
&\max_{\bm{\beta},\mathbf{I},\mathbf{r}} ~
f_0\left(\mathbf{r}\right)
- \mathcal{L}^{(1)}\left(\bm{\alpha},\bm{\beta},\bm{\lambda}\right) - \mathcal{L}^{(2)}\left(\bm{\alpha},\bm{\beta},\bm{\widetilde{\lambda}}\right) \tag{\ref{Pbeta}{a}}
\\{\mathrm{s.t.}}~ & \eqref{constraint_var2}, \eqref{constraint_intf2}, \eqref{constraint_rate2}, \eqref{constraint_rubSCA}, \eqref{constraint_rlbSCA}. \tag{\ref{Pbeta}{d}}
\vspace{-1.4em}
\end{align*}
\end{subequations}
\vspace{-1cm}

Since \eqref{Pbeta} is a convex optimization problem, it can be easily solved by the interior point method using the standard convex optimization tool, such as CVX \cite{CVX}.

Furthermore, at each iteration $t$ the dual variables $\bm{\lambda}$ and $\bm{\widetilde{\lambda}}$ can be updated by
\vspace{-0.5em}
\begin{equation}\label{Lambda1}
\bm{\lambda}^{(t+1)}=\bm{\lambda}^{(t)}+\frac{1}{\rho}\left(\bm{\beta}^{(t)}+\mathbf{\bm{\alpha}}^{(t)}-\mathbf{1}_{K\times K}\right),
\vspace{-0.5em}
\end{equation}
\begin{equation}\label{Lambda2}
\widetilde{\lambda}_{ik}^{(t+1)}=\widetilde{\lambda}_{ik}^{(t)}+\frac{1}{\rho}\beta_{ik}^{(t)}\alpha_{ik}^{(t)}, ~ \forall i,k \in \mathcal{K}.
\vspace{-0.5em}
\end{equation}

\vspace{-0.5em}
\begin{algorithm}[H]
\caption {ADMM-SCA Algorithm for Solving $\mathcal{P}_{1}$}
\begin{algorithmic}[1]
\STATE Initialize the accuracy tolerance $\epsilon_{\mathrm{ADMM}} > 0$ and the maximum iteration number $T_{\mathrm{ADMM}}^{\max}$.
\STATE Initialize $\left\{\bm{\alpha},\bm{\beta},\mathbf{W},\mathbf{I},\mathbf{S},\mathbf{r},\bm{\lambda},\bm{\widetilde{\lambda}}\right\}$ with a feasible point, and initialize $\rho > 0$.
\STATE Set the iteration number as $t=0$.
\REPEAT
    \STATE By fixing $\left\{\bm{\beta},\overline{\bm{\alpha}},\overline{\mathbf{W}},\overline{\mathbf{I}},\overline{\mathbf{r}},\bm{\lambda},\widetilde{\bm{\lambda}}\right\}$, update the SIC operations $\bm{\alpha}$ and the transmit beamforming $\mathbf{W}$ by solving problem \eqref{Palpha}.
    \STATE By fixing $\left\{\bm{\alpha},\mathbf{W},\mathbf{S},\overline{\bm{\alpha}},\overline{\mathbf{I}},\overline{\mathbf{r}},\bm{\lambda},\widetilde{\bm{\lambda}}\right\}$, update the variables $\bm{\beta}$ by solving problem \eqref{Pbeta}.
    \STATE Update dual variables $\bm{\lambda}$ and $\bm{\widetilde{\lambda}}$ using \eqref{Lambda1} and \eqref{Lambda2}, respectively.
\UNTIL{$t = T_{\mathrm{ADMM}}^{\max}$ or the difference of successive objective values satisfies  $\left|f_0\left(\bm{r}^{t}\right)-f_0\left(\bm{r}^{t-1}\right)\right|^2 \le \epsilon_{\mathrm{ADMM}}$.}
\ENSURE The SIC operations $\bm{\alpha}^{*}$, transmit beamforming $\mathbf{W}^{*}$, and the optimal value.
\end{algorithmic}
\label{alg:ADMMSCA}
\end{algorithm}
% \vspace{-0.5cm}

The overall ADMM-SCA algorithm can be summarized as \textbf{Algorithm \ref{alg:ADMMSCA}}.
The computational complexity of solving convex subproblems \eqref{Palpha} and \eqref{Pbeta} via the interior point method can be respectively given by $\mathcal{O}\left(\left(4K^{2}+MK\right)^{3.5}\right)$ and $\mathcal{O}\left(\left(3K^{2}\right)^{3.5}\right)$ \cite{ConvexOpt}.
Therefore, the computational complexity of \textbf{Algorithm \ref{alg:ADMMSCA}} is $\mathcal{O}\left(T\left(4K^{2}+MK\right)^{3.5}+\left(3K^{2}\right)^{3.5}\right)$, where $T$ denotes the number of iterations for reaching convergence.
According to the analyses in \cite{SCA_converegence} and \cite{ADMM_convergence}, the ADMM-SCA algorithm can converge to a feasible and stationary solution of problem $\mathcal{P}_1$ with polynomial time complexity.
However, the ADMM framework generally suffers from slow convergence and requires a high computational complexity.
Moreover, the obtained discrete variables $\bm{\alpha}$ and $\bm{\beta}$ are usually highly sensitive to the initialized parameters, which thus significantly impact the resulting performances.
Hence, we randomly initialize $\mathbf{W}$, and test $N^{\mathrm{ini}}$ groups of initialized parameters for $\left\{\bm{\beta},\bm{\alpha},\bm{\mathbf{S}},\bm{I}\right\}$ to empirically choose the initialization points in different communication regimes.

\vspace{-0.1em}
\section{Low-Complexity Matching-SCA Based Solution}
Although the ADMM-SCA algorithm achieves monotonic convergence to a desirable suboptimal solution, it may need a large number of iterations for convergence and require high computational complexity when user number increases. Moreover, the achieved performance is typically highly sensitive to the initialized parameters due to the discrete optimization.
To overcome these shortcomings, in this section we further propose a novel low-complexity and efficient strategy, which solves the non-convex NP-hard MINLP based on the matching game theory and the inexact SCA method.

\vspace{-0.8em}
\subsection{Many-To-Many SIC Matching Problem}
Firstly, we model the cluster-free SIC optimization as a dynamic two-sided matching game among the connected users.
We define two virtual user sets $\mathcal{U}$ and $\mathcal{V}$ with logically disjoint entries, where $\mathcal{U}$ consists of the users that execute SIC to cancel the interference imposed by users from $\mathcal{V}$.
Without loss of generality, we define $\mathcal{U} = \mathcal{V} = \mathcal{K}$.
If user $u\in\mathcal{U}$ is scheduled to carry out SIC to eliminate interference from user $v\in\mathcal{V}$, we say user $u$ and user $v$ are matched to each other, which is denoted by $(u,v)$.

The SIC operations can be formulated as a matching problem, which yields the following definitions and remarks.

\vspace{-0.6em}
\begin{definition}[Many-to-Many Matching]
A man-to-many matching $\mu$ is a function from set $\mathcal{U}\cup\mathcal{V}$ to the set of all subsets of $\mathcal{U}\cup\mathcal{V}$, such that
\begin{itemize}[\itemindent=1em]
  \item[(i)]$\mu(u)\subset\mathcal{V}$ and $\left|\mu(u)\right|\le N_{u}$, $\forall u\in\mathcal{U}$;
  \item[(ii)]$\mu(v)\subset\mathcal{U}$ and $\left|\mu(v)\right|\le N_{v}$, $\forall v\in\mathcal{V}$;
  \item[(iii)]$\mu(u)\subset\mathcal{V}$ if and only if $\mu(v)\subset\mathcal{U}$;
  \item[(iv)]$u\in\mu(v)$ if and only if $v\in\mu(u)$.
\end{itemize}
\vspace{-0.8em}
\end{definition}
In the above definition, condition (i) means that each user $u\in\mathcal{U}$ can carry out SIC for a subset of users in $\mathcal{V}$, and the cardinality of $\mu(u)$ cannot exceed $N_u$.
Condition (ii) indicates that the interference from each user $v\in\mathcal{V}$ can be eliminated with SIC by at most $N_v$ users from  $\mathcal{U}$.
Condition (iii) represents that the mapping of user $u\in\mathcal{U}$ is the subset of $\mathcal{V}$, and vice versa.
Condition (iv) implies that when $u\in\mathcal{U}$ matches with $v\in\mathcal{V}$, $v$ matches with $u$ as well.
Without loss of generality, we set $N_{u}=N_{v}=K$ here. Thus, both the cluster-based and beamformer-based NOMA approaches can be included as special cases of the proposed strategy.

During the matching process, each user $u\in\mathcal{U}$ and $v\in\mathcal{V}$ have their individual preference lists.
Given a matching $\mu$, we formulate the utility function $U_{k}^{\mu}$, i.e., the preference value of each user $k\in\mathcal{U}\cup\mathcal{V}$ over matching $\mu$, as its achievable data rate while fixing the matching states of the other users, which can be defined as
\vspace{-0.4em}
\begin{equation}\label{Utility_u}
U_{k}^{\mu} \!=\! R_{k}^{\mu}\left(\boldsymbol{\alpha}^{\mu}\!,\mathbf{W}^{\mu}\right)
\!=\!
\min \bigg\{R_{k\rightarrow k}\left(\boldsymbol{\alpha}^{\mu},\mathbf{W}^{\mu}\right),\!
\left\{\frac{1}{\alpha_{ik}^{\mu}}R_{i\rightarrow k}\left(\boldsymbol{\alpha}^{\mu},\mathbf{W}^{\mu}\right)\right\}_{i\ne k}\bigg\}, \forall k\in\mathcal{U}\cup\mathcal{V}.
\vspace{-0.4em}
\end{equation}
Here, \eqref{Utility_u} returns the achievable data rate of user $u$ as implied by the SIC constraint $\frac{1}{\alpha_{ik}} R_{k\rightarrow k}\le R_{i\rightarrow k}$ from \eqref{constraint_SIC}, where $R_{k}^{\mu}$, $\boldsymbol{\alpha}^{\mu}$, and $\mathbf{W}^{\mu}$ denotes the data rate of user $k$, the SIC operations, and the beamforming coefficients corresponding to the matching $\mu$, respectively.
Therefore, the total utility over the matching $\mu$ can be expressed as $U^{\mu} = \sum\limits_{k\in\mathcal{K}} U_k^{\mu}$.

\vspace{-0.25em}
\begin{lemma}\label{remark:matching_property}
The formulated two-sided many-to-many matching problem has the properties of externality and non-substitutability.
\begin{proof}
The properties can be demonstrated as follows.
\textbf{i) Externality}: Owing to the feature of the multi-antenna NOMA system, the interference suffered by each user $k\in\mathcal{U}\cup\mathcal{V}$ varies with matching states of the other users.
Therefore, the achievable data rate and preference of each user $k$ also depend on other users, and each user should take into account the internal relationship of the other users when determines its matching state. This renders the externality of the SIC matching problem.
\textbf{ii) Non-substitutability}: Given two virtual user sets $\mathcal{U}$ and $\mathcal{V}$, each user $u\in\mathcal{U}$ prefers to match with a subset $\mathcal{V}_{u}$ of $\mathcal{V}$, which is defined as the choice of $u$ in $\mathcal{V}$, denoted by $\mathcal{C}_{u}\left(\mathcal{V}\right) = \mathcal{V}_{u}$.
Here, user $u$ prefers $\mathcal{V}_{u}$ to any subset of $\mathcal{V}$, which can be denoted as $\mathcal{V}_{u} \succ_{u} \mathcal{V}'$, $\forall \mathcal{V}' \subset \mathcal{V}, ~ \mathcal{V}' \ne \mathcal{V}_{u}$.
The preference of $u$ over sets of $\mathcal{V}$ possesses substitutability property if and only if $v\in\mathcal{C}_{u}\left(\mathcal{V}\right)$ and $v\in\mathcal{C}_{u}\left(\mathcal{V} \setminus \{v'\}\right)$, $\forall v, v' \in \mathcal{V}$, which means that the choice of $u$ in $\mathcal{V}$ will not be affected even if one matched user in $\mathcal{V}$ is excluded.
However, since the optimal beamforming and inter-user interference varies with the SIC operations, the achievable data rate may change under different user matching.
Thus, the formulated matching based SIC operation problem lacks the substitutability property.
\end{proof}
\end{lemma}
\vspace{-0.8em}

To address the externality and ensure exchange stability, we first introduce the following matching swap operation as defined by conventional matching theory \cite{MatchingExternality_Zhao,MatchingExternality_Ni}
\vspace{-0.8em}
\begin{equation}\label{Swap}
\mu_{uv}^{u'v'} = \left\{ \mu \setminus \left\{(u,v),(u',v')\right\} \cup \left\{(u',v), (v,u')\right\} \right\},
\vspace{-0.9em}
\end{equation}
which means that user $u$ and $u'$ exchange their matched users $v$ and $v'$ while keeping all other users' matching states unchanged.
Here, we also consider the swap operation over ``holes", i.e., the empty set $\varnothing$ that does not contain any users.
Specifically, the matching state of user pair $(u,v)$ can be transferred from matched into unmatched after the swap operation $\mu_{uv}^{\varnothing\varnothing}$.
Furthermore, the state of the user pair $(u,v)$ can shift from unmatched to matched based on $\mu_{u\varnothing}^{\varnothing v}$.
Based on the matching swap operation, the swap-blocking pair can be defined as follows.

\vspace{-0.5em}
\begin{definition}[Swap-Blocking Pair]\label{SwapRule}
For two users $u$ and $u'$, $(u,u')$ is a swap-blocking pair in matching $\mu$ if and only if
\begin{itemize}[\itemindent=1em]
  \item[(i)] $\forall k \in \{u,u',v,v'\}$, $U_{k}\left(\mu_{uv}^{u'v'}\right) \ge U_{k}(\mu)$;
  \item[(ii)] $\exists k \in \{u,u',v,v'\}$, such that $U_{k}\left(\mu_{u}^{u'}\right) > U_{k}(\mu)$;
\end{itemize}
where $U_k(\mu)$ denotes the utility of a user $k\in\mathcal{U}\cup\mathcal{V}$ over matching $\mu$.
\vspace{-0.8em}
\end{definition}

\vspace{-0.3cm}
\subsection{Extended Many-To-Many Matching for SIC Operations}

In the proposed framework, the SIC operations $\bm{\alpha}$ determine both the user matching and the decoding order, as implied by \eqref{DecodingOrder}.
However, the traditional swap operation in \eqref{Swap} only optimizes the user matching, and cannot dynamically and jointly optimize the NOMA SIC decoding order.
To address this problem, we extend the traditional matching algorithm to efficiently optimize $\bm{\alpha}$ for the generalized cluster-free multi-antenna NOMA.

In contrast to the traditional matching model, the matched users $(u,v)$ and $(v,u)$ in the formulated matching game have completely different physical meanings, which lead to swaps of SIC  decoding orders.
Therefore, we propose a decoding order swap operation, which enables the exchange of SIC decoding order between two matched users $u$ and $v$.
The decoding order swap operation $\widetilde{\mu}_{uv}^{u'v'}$ can be defined as follows:
\vspace{-0.9em}
\begin{equation}\label{ESwap}
\widetilde{\mu}_{uv}^{u'v'} = \left\{ \mu \setminus \left\{(u,v)\right\} \cup \left\{(u',v')\right\} \right\},
~ u' = v, ~ v' = u.
\vspace{-1em}
\end{equation}

Combining the conventional user matching swap operation \eqref{Swap} and the decoding order swap operation \eqref{ESwap}, we further present the following concept of enhanced swap-blocking pair to determine the swap rule.
Different from conventional matching swap rule that ensures the utility increments of individual users (see \textbf{Definition \ref{SwapRule}}), we aim at improving the sum utility $U(\mu)$ of all users in matching $\mu$ to maximize the sum rate.

\vspace{-0.2em}
\begin{definition}[Enhanced Swap-Blocking Pair]\label{definition_ESBP}
Given a matching $\mu$, for user $u\in\mathcal{U}$ and user $u'\in\mathcal{U}$, we define $(u,u')$ as an enhanced swap-blocking pair, if one of the following conditions can be satisfied
\begin{itemize}[\itemindent=1em]
  \item[(i)] $\exists v\in\mu(u), v'\in\mu(u')$, such that $U\big(\mu_{uv}^{u'v'}\big) > U(\mu)$ and $\alpha_{v'u}=\alpha_{vu'}=0$;
  \item[(ii)] for $v=u',~v'=u$, we can obtain that $v = \mu(u)$ and $U\big(\widetilde{\mu}_{uv}^{u'v'}\big) > U(\mu)$.
\end{itemize}
\vspace{-0.2em}
\end{definition}

The above condition (i) indicates that after a conventional matching swap \eqref{Swap}, the overall utility should be increased.
Moreover, to ensure that the matching swap from $\left\{(u,v), (u',v')\right\}$ to $\left\{(u,v'), (u',v)\right\}$ is feasible, we should always guarantee $\alpha_{uv'}+\alpha_{v'u}\le 1$ and $\alpha_{vu'}+\alpha_{u'v}\le 1$, which leads to the constraint $\alpha_{v'u}=\alpha_{vu'}=0$ in condition (i).
Condition (ii) implies that  after a decoding order swap operation \eqref{ESwap}, the overall utility should be improved.

Note that \textbf{Definition \ref{definition_ESBP}} determines the swap rule of the formulated matching.
To be more specific, if there exists an enhanced swap-blocking pair satisfying any of the above conditions (i) and (ii), then the matching is not stable and convergent, and the corresponding matching $\mu_{uv}^{u'v'}$ or $\tilde{\mu}_{uv}^{u'v'}$ would be ``approved".
Therefore, by extending the concept of exchange stability in conventional matching \cite{Stability_Bodine}, we can define the enhanced exchange stability as follows.

\vspace{-0.5em}
\begin{definition}[Enhanced Exchange-Stable Matching]
The two-sided matching $\mu$ is an enhanced exchange-stable matching if and only if there dose not exist an enhanced swap-blocking pair.
\end{definition}

\vspace{-1.4em}
\subsection{Matching-SCA Based Joint Optimization}
Based on the proposed extended matching, we further develop a dual-loop iterative algorithm to jointly optimize the SIC operations and the transmit beamforming.
In the outer loop, the matching state $\mu$ of SIC operations are updated by the extended many-to-many matching while fixing the transmit beamforming.
In the inner loop, given the SIC operations, the transmit beamforming $\mathbf{W}$ is sequentially optimized via an SCA process. 

The inner-loop transmit beamforming optimization can be illustrated as follows. Given the current matching state $\mu$ and the corresponding SIC operation variables $\bm{\alpha}^{\mu}$, the beamforming $\mathbf{W}$ can be optimized by invoking the SCA method, as analysed in Section \ref{Section_ADMM_SCA}. %\textbf{Lemma \ref{Lemma_SCA}}.
By fixing $\bm{\alpha}^{\mu}$, $\mathbf{W}$ can be optimized by sequentially solving the following convex problem
\vspace{-1em}

\vspace{-1.2em}
\begin{subequations}\label{PBF}
\begin{align*}
& \max_{\mathbf{W},\mathbf{S},\mathbf{I},\mathbf{r}} ~
f_0\left(\mathbf{r}\right) \tag{\ref{PBF}{a}}
\\{\mathrm{s.t.}}~ &
r_{ik}+\log_{2}\left(\overline{I}_{ik}\right)+\frac{1}{\ln2}\frac{1}{\overline{I}_{ik}}\left(I_{ik}-\overline{I}_{ik}\right)
\le \log_{2}\left(I_{ik}+S_{ik}\right), ~ \forall i,k\in\mathcal{K}, \label{constraint_rubBF} \tag{\ref{PBF}{b}}
\\&
r_{ik}\ge \alpha_{ik}^{\mu} r_{kk}, ~ \forall i,k\in\mathcal{K}, ~ i\ne k, \label{constraint_rlbBF} \tag{\ref{PBF}{c}}
\\&
S_{ik} +\left|\mathbf{h}_{i}^{H}\mathbf{\overline{w}}_{k}\right|^{2}
\le 2\mathscr{R}\left(\overline{\mathbf{w}}_{k}^{H}\widetilde{\mathbf{H}}_{i}\mathbf{w}_{k}\right), ~\forall i,k\in\mathcal{K},  \label{constraint_signalBF} \tag{\ref{PBF}{d}}
\\&
\widetilde{\mathrm{Intf}}_{i\rightarrow k}\left(1-\bm{\alpha}^{\mu},\mathbf{W}\right) \le I_{ik}, ~ \forall i,k \in\mathcal{K}, \tag{\ref{PBF}{e}}
\\&
\eqref{constraint_rate2}, \eqref{constraint_power2}.  \label{constraint_varBF} \tag{\ref{PBF}{f}}
\end{align*}
\vspace{-0.5cm}
\end{subequations}

\begin{algorithm}[!h]
\caption{Matching-SCA Algorithm for Solving $\mathcal{P}_0$}
\begin{algorithmic}[1]
\STATE Initialize the algorithm accuracy $\epsilon_{\mathrm{MSCA}}>0$. Set the maximal outer and inner loop iteration number as $T_{\mathrm{MSCA}}^{\max}=20$ and $L^{\max}=3$.
\STATE Initialize the matching states as $\bm{\alpha} = \mathbf{I}_{K\times K}$ with all users unmatched, and initialize $\mathbf{W}$ with a feasible point.
\STATE Set the outer loop iteration number $t=0$.
\REPEAT
    \STATE \textbf{\textit{// (Inner loop) SCA-based transmit beamforming optimization}}
    \STATE Set the inner loop iteration number $l=0$.
    \REPEAT
        \STATE Update $l\leftarrow l+1$.
        \STATE Update $\mathbf{W},\mathbf{I},\mathbf{S},\mathbf{r}$ by solving \eqref{PBF}.
    \UNTIL{$l\ge L^{\max}$}.
    \STATE \textbf{\textit{// Matching-based SIC operation optimization}}
    \STATE For every user $u\in\mathcal{U}$, search for another $u'\in\mathcal{U}\cup\{\varnothing\}$.
    \IF{$(u,u')$ forms an enhanced swap-blocking pair satisfying condition (i) in \textbf{Definition \ref{definition_ESBP}}}
        \STATE Update $\mu \leftarrow \mu_{uv}^{u'v'}$.
    \ENDIF
    \IF{$(u,u')$ forms an enhanced swap-blocking pair satisfying condition (ii) in \textbf{Definition \ref{definition_ESBP}}}
        \STATE Update $\mu \leftarrow  \widetilde{\mu}_{uv}^{u'v'}$.
    \ENDIF
    \STATE Update $\mathbf{S}$ and $\mathbf{I}$ based on current $\mathbf{W}$ and $\bm{\alpha}^{\mu}$ using \eqref{update_S} and \eqref{update_I}.
    \STATE Update $t\leftarrow t+1$.
\UNTIL{the difference of successive objective values satisfy
$\left|f_0\left(\bm{r}^{t}\right)-f_0\left(\bm{r}^{t-1}\right)\right|^2 \le \epsilon_{\mathrm{MSCA}}$ or $t\ge T_{\mathrm{MSCA}}^{\max}$.}
\ENSURE Matching $\mu$, SIC operation $\bm{\alpha}^{\mu}$, transmit beamforming $\mathbf{W}$, and the optimal value.
\end{algorithmic}
\label{alg:MatchingSCA}
\end{algorithm}
% \vspace{-0.3cm}

The developed joint optimization algorithm, namely Matching-SCA, can be summarized as \textbf{Algorithm \ref{alg:MatchingSCA}}.
Firstly, the initialized matching states of all users is set as unmatched, i.e., $\bm{\alpha} = \mathbf{I}_{K\times K}$.
Moreover, the transmit beamforming $\mathbf{W}$ is randomly initialized with a feasible point.
The iterative procedure exploits a dual-loop structure.
Specifically, given the current matching states $\mu$, the beamforming coefficients $\mathbf{W}$ are sequentially optimized via SCA in the inner loop to an inexact solution that is not required to be locally converged (step 7-step 12).
Thereafter, by fixing the transmit beamforming $\mathbf{W}$, we further perform an enhanced swap-matching process in the outer loop (step 13-step 20).
By searching enhanced swap-blocking pairs, preferable matching swaps and decoding order swaps can be obtained to improve the sum rate.
Based on the resulting matching $\bm{\alpha}^{\mu}$ and $\mathbf{W}$, we further update the auxiliary variables $\mathbf{I}$ and $\mathbf{S}$ as
\vspace{-0.5em}
\begin{equation}\label{update_S}
S_{ik}^{\mu} = \left|\mathbf{h}_{i}^H\mathbf{w}_{k}\right|^2, ~ \forall i,k\in\mathcal{K},
\vspace{-0.5em}
\end{equation}
\begin{equation}\label{update_I}
I_{ik}^{\mu} = \mathrm{Intf}_{i\rightarrow k} \left(\bm{\alpha}^{\mu},\mathbf{W}\right),
~ \forall i,k\in\mathcal{K}.
\vspace{-0.5em}
\end{equation}
The above process is repeated until the termination criterion is reached.

\vspace{-0.5em}
\subsection{Theoretical Analysis}
\vspace{-0.2em}
The properties of the proposed Matching-SCA algorithm with regard to the stability, convergence, and optimality can be theoretically analysed as follows.

\vspace{-0.5em}
\begin{proposition}[Stability]\label{proposition:Stability}
The proposed \textbf{Algorithm \ref{alg:MatchingSCA}} eventually reaches an enhanced two-sided exchange-stable matching.
\begin{proof}
We prove this proposition by contradiction.
Assume that there exists an enhanced blocking pair $(u,u')$ in the resulting $\mu^{*}$, which satisfies condition (i) or condition (ii) in \textbf{Definition \ref{definition_ESBP}}.
According to step 14-step 20 in \textbf{Algorithm \ref{alg:MatchingSCA}}, the proposed algorithm will continue swap until no enhanced blocking pair satisfying the swap conditions exists in the current matching.
That is to say, $\mu^{*}$ should not be the resulting matching, which contradicts the initial assumption.
Therefore, it can be concluded that an enhanced exchange stability can be achieved by the proposed Matching-SCA Algorithm \ref{alg:MatchingSCA} eventually.
This completes the proof.
\end{proof}
\end{proposition}

\vspace{-1em}
\begin{proposition}[Convergence]\label{proposition:Convergence}
\textbf{Algorithm \ref{alg:MatchingSCA}} converges to an enhanced two-sided exchange-stable matching within limited swap operations.
\begin{proof}
Given a matching function $\mu$ for the SIC operation problem, assume that $(u,u')$ is an enhanced swap-blocking pair.
Based on \textbf{Definition \ref{definition_ESBP}}, there are two cases for $(u,u')$:
i) for users $v=\mu(u)$ and $v'=\mu(u')$ satisfying $\alpha_{v'u}=\alpha_{vu'}=0$, a conventional swap matching is ``approved", i.e., $U\big(\mu_{uv}^{u'v'}\big) > U(\mu)$;
ii) for $v=u'$ and $v'=u$ satisfying $\mu(u)=v$, a decoding order swap matching is ``approved", i.e., $U\big(\widetilde{\mu}_{uv}^{u'v'}\big) > U(\mu)$.
Since the utilities of all users are non-decreasing for both cases, the sum rate is non-decreasing after each swap operations.
Furthermore, owing to the limited transmit power and the inter-user interference, it can be observed that the achievable rate is upper bounded in practice.
Therefore, the number of swap operations is limited for every swap-matching process in Algorithm \ref{alg:MatchingSCA}.
Based on \textbf{Proposition \ref{proposition:Stability}}, the proposed algorithm eventually converges to an enhanced two-sided exchange-stable matching when neither matching swaps nor decoding order swaps can further improve the total utility.
This completes the proof.
\end{proof}
\vspace{-0.2em}
\end{proposition}

\vspace{-1em}
\begin{theorem}[Local Optimality]\label{proposition:Optimality}
\textbf{Algorithm \ref{alg:MatchingSCA}} converges to a locally optimal matching and beamforming solution.
\begin{proof}
See Appendix \ref{appendix:proof_optimality}.
\end{proof}
\end{theorem}

\vspace{-0.6cm}
\section{Numerical Results}
In this section, we present simulation results to verify the effectiveness of the proposed generalized cluster-free multi-antenna NOMA framework and algorithms.
We consider one BS which is equipped with $M=4$ antennas to serve $K = \{3,4,...,11\}$ users randomly deployed in its coverage range in the downlink transmissions.
The signal-to-noise-ratio (SNR) is set as $20$ dB, and the maximum transmission power of the BS is $27$ dBm.
To characterize the channel spatial correlations of users, we model the channel gains $\mathbf{H}$ as \cite{ChannelCorr}
\vspace{-0.5em}
\begin{equation}\label{ChannelGain}
\mathbf{H} = \widetilde{\mathbf{H}}\mathbf{R}_{\mathbf{H}}^{1/2},
\vspace{-0.5em}
\end{equation}
where $\widetilde{\mathbf{H}}$ is the normalized Rayleigh fading matrix which satisfies $\mathbb{E}\left[\widetilde{\mathbf{H}}^H\widetilde{\mathbf{H}}\right]=\mathbf{I}$. $\mathbf{R}_{\mathbf{H}}$ denotes the covariance of $\mathbf{H}$, where the $(i,j)$-th element signifies the channel spatial correlation of user $i$ and user $j$.
For each channel realization, $\mathbf{R}_{\mathbf{H}}$ can be mathematically formulated as
\vspace{-0.5em}
\begin{equation}
\mathbf{R}_{\mathbf{H}}=\left[\begin{array}{ccccc}
1 & c & c^{2} & ... & c^{K}\\
c^{H} & 1 & c & ... & c^{K-1}\\
\left(c^{2}\right)^{H} & c^{H} & 1 & ... & c^{K-2}\\
... & ... & ... & ... & ...\\
\left(c^{K}\right)^{H} & \left(c^{K-1}\right)^{H} & \left(c^{K-2}\right)^{H} & ... & 1
\end{array}\right],
\vspace{-0.3em}
\end{equation}
where $c = corr\times e^{j\phi}$ with $\phi$ being the randomly generated phase within $[0,2\pi]$ and $corr$ controlling the mean channel correlation.
For different communication regimes, $N^{\mathrm{ini}}=20$ groups of initialization parameters are tested to empirically select the initialization points for the ADMM-SCA algorithm.

\vspace{-0.6em}
\subsection{Convergence Behavior}

\begin{figure}[!t]
\vspace{-1.8em}
	\centering
	\subfloat[Convergence of the proposed algorithms.]{\centering \scalebox{0.41}{\includegraphics{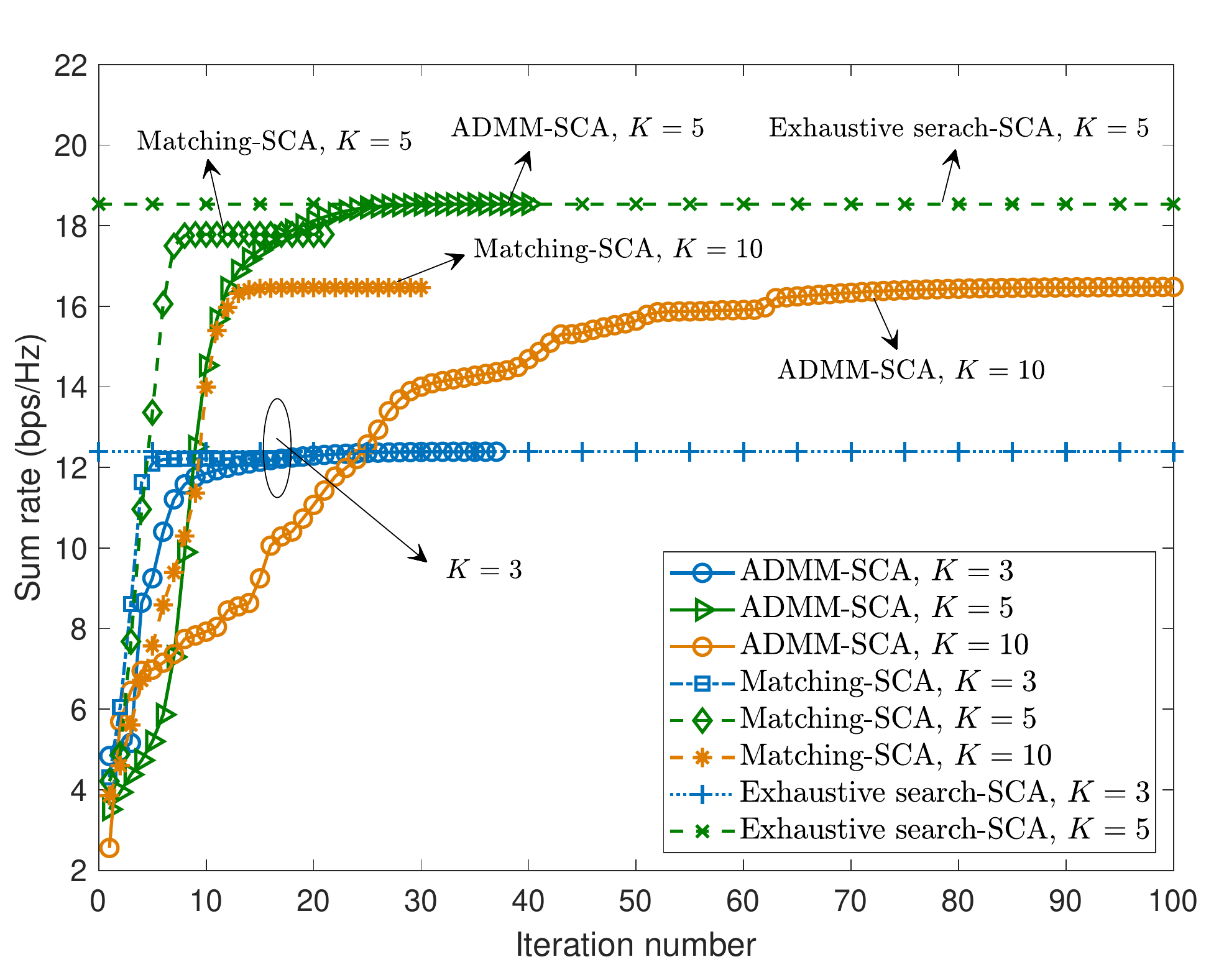}} }
	\subfloat[Comparisons of the matching-based algorithms. $K=10$.]{\centering \scalebox{0.41}{\includegraphics{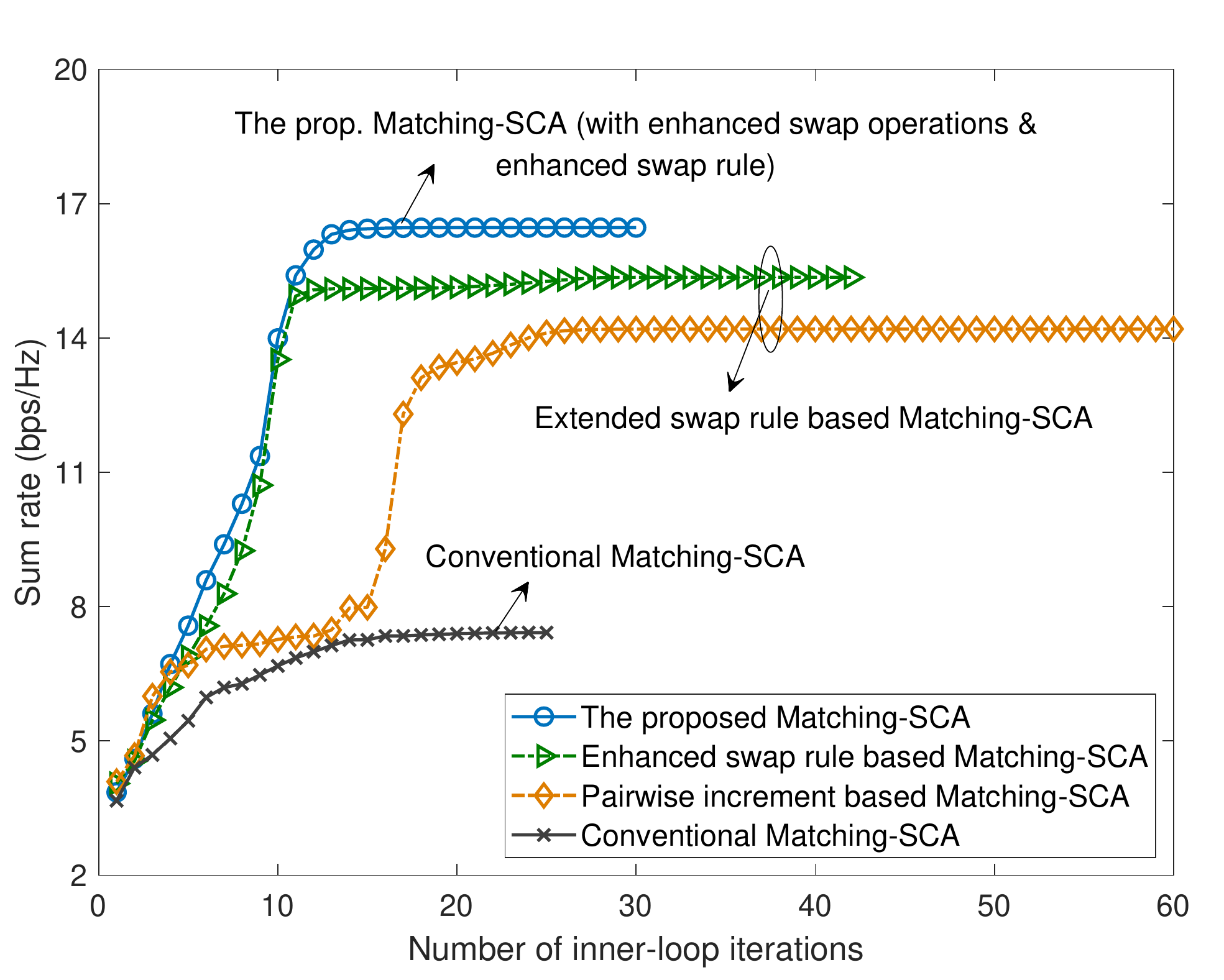}} }	
	\caption{Convergence behaviors of different algorithms.}\label{fig_conv}
\vspace{-1em}
\end{figure}

Fig. \ref{fig_conv} presents the convergence behaviors of different algorithms, where we set $corr=0.9$.
In Fig. \ref{fig_conv}(a), we first compare the convergence behaviors under different user numbers.
Here, we introduce the \textit{exhaustive search-SCA} algorithm as a benchmark, which solves beamforming coefficients $\mathbf{W}$ by SCA while obtaining the globally optimal $\bm{\alpha}$ by exhaustively searching all possible combinations.
As shown in Fig. \ref{fig_conv}(a), if initialized parameters are well tuned, the ADMM-SCA algorithm can achieve the near-optimal SIC operations indicated by \textit{exhaustive search-SCA}.
However, ADMM-SCA incurs high computational complexity and takes more than $60$ iterations for convergence when the number of users increases.
Moreover, the Matching-SCA algorithm yields close performances to ADMM-SCA while achieving much faster convergence, which can converge within $30$ inner-loop iterations even in the severely overloaded system.
It is worth pointing out that Matching-SCA may outperform ADMM-SCA especially in the overloaded systems. This is because the binary SIC operation variables $\bm{\alpha}$ obtained by ADMM-SCA is highly sensitive to initialized parameters, but Matching-SCA can alleviate the over-dependence on the parameter initialization.

In Fig. \ref{fig_conv}(b), we further compare the proposed Matching-SCA algorithm with conventional matching-based algorithms.
Here, three baseline matching-based algorithms are considered:
\begin{itemize}
    \item \textbf{Conventional Matching-SCA}: where the conventional matching swap operations and the conventional swap rule are utilized, as presented in \textbf{Definition \ref{SwapRule}}.
    \item \textbf{Pairwise increment based Matching-SCA}: where the conventional matching swap operation (without the swaps of SIC decoding order) is utilized, while the swap rule is modified into ensuring the sum utility increment of the involved user pairs during swaps.
    \item \textbf{Enhanced swap rule based Matching-SCA}: where the conventional matching swap operation (without the swaps of SIC decoding order) is utilized, while the enhanced swap rule as described in condition (i) in \textbf{Definition \ref{definition_ESBP}} is adopted, which ensures the sum utility increment of all multiplexed users.
\end{itemize}
From Fig. \ref{fig_conv}(b), the conventional Matching-SCA algorithm leads to the worst sum rate since it cannot optimize the SIC decoding order and only focuses on users' individual utilities, which typically lacks the local optimality guarantee at the convergence.
Furthermore, the enhanced swap rule based Matching-SCA algorithm outperforms the pairwise increment based Matching-SCA algorithm, since it takes into account the total utility of all users.
The proposed Matching-SCA algorithm achieves the highest performance, which validates the effectiveness of the enhanced swap operations and enhanced swap rule as proposed in \textbf{Definition \ref{definition_ESBP}}.

\vspace{-0.5em}
\subsection{Performance Comparisons}
To demonstrate the performance of the proposed generalized cluster-free framework as analysed in \textbf{Remark \ref{remark_generalization}}, we consider four baseline approaches, namely cluster-based NOMA (CB-NOMA), enhanced cluster-based NOMA (enhanced CB-NOMA), beamformer-based NOMA (BB-NOMA), and SDMA.
Specifically, the enhanced CB-NOMA is an improved variant of the conventional CB-NOMA.
While the CB-NOMA sharing a single beamforming vector among each cluster, the enhanced CB-NOMA exploits dedicated beamforming vectors for each user to achieve possibly better spatial multiplexing.
Both CB-NOMA and enhanced CB-NOMA configure SIC operations by performing user clustering according to users' channel correlations \cite{UserClustering_Dai}.
Without loss of the generality, the beamforming vectors of the enhanced CB-NOMA, BB-NOMA, and SDMA are solved by sequentially optimizing problem \eqref{PBF} via SCA.
Moreover, we accordingly optimize the cluster-specific beamforming and the power allocation of CB-NOMA by modifying problem \eqref{PBF} and invoke the SCA and alternating optimization.
We obtain the following simulation results by averaging over $100$ channel realizations.

\vspace{-0em}
\begin{figure}[!h]
\vspace{-0.2em}
	\centering
	\subfloat[Sum rate.]{
        \centering \scalebox{0.41}{\includegraphics{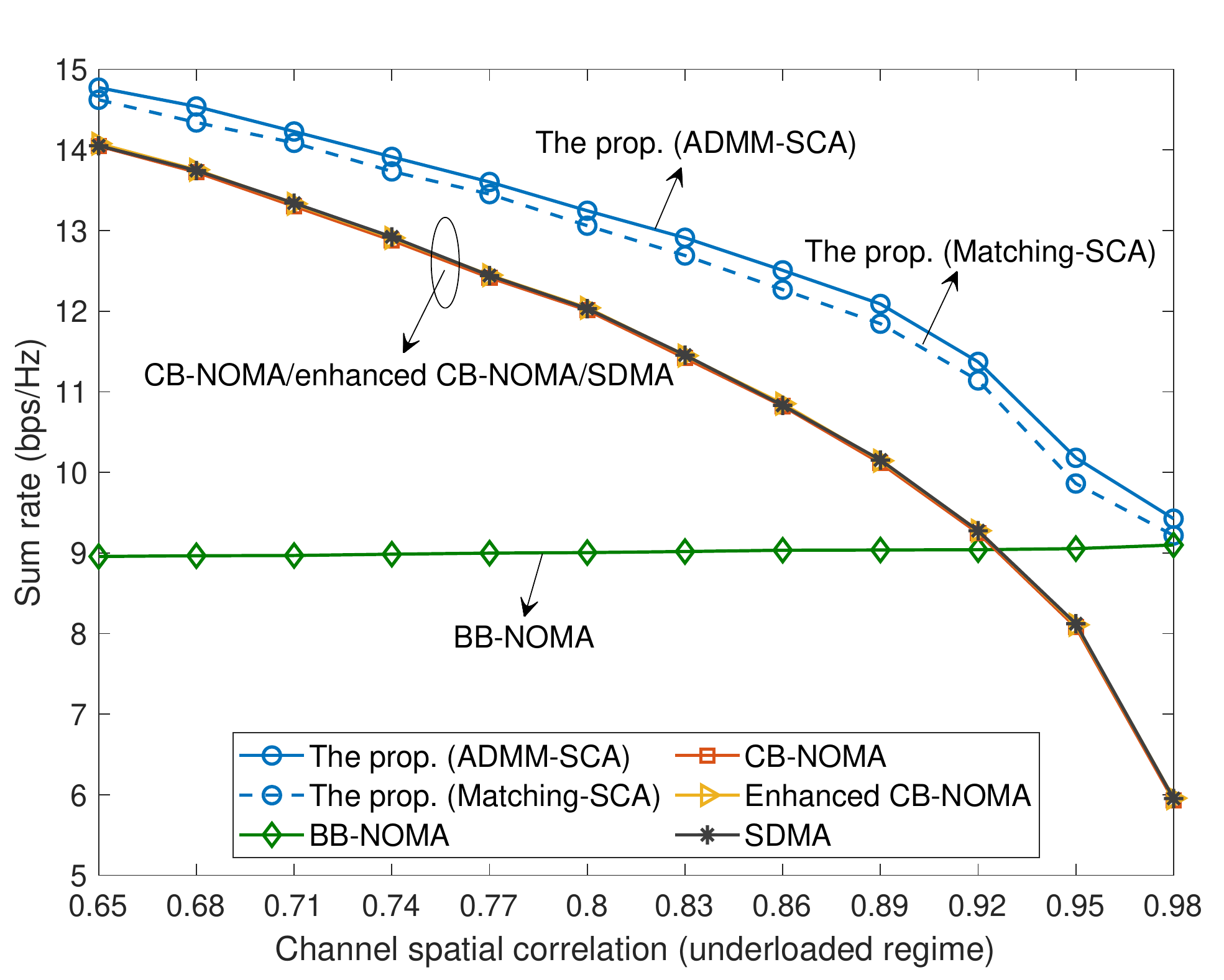}}
    }
	\subfloat[SIC decoding complexity.]{
        \centering \scalebox{0.41} {\includegraphics{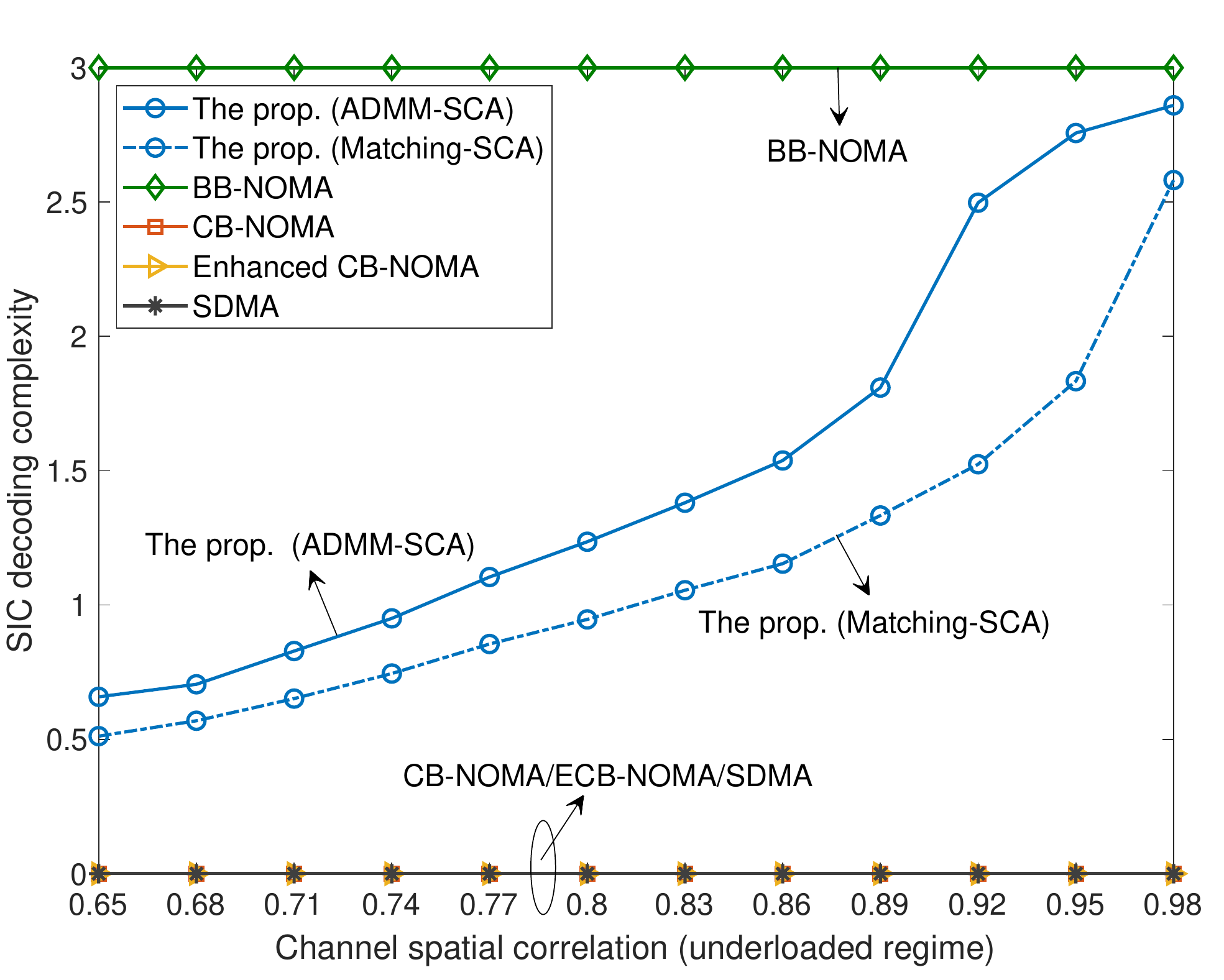}}
    }
	\caption{Performance comparisons under various channel correlations $corr$ in the underloaded regime. $M=4$, $K=3$.}
	\label{fig_corr_ul}
\vspace{-0.5em}
\end{figure}

Fig. \ref{fig_corr_ul} demonstrates the performances of different methods under different channel correlations in the underloaded regime, where $M=4$ and $K=3$.
From Fig. \ref{fig_corr_ul}(a), it can be observed that the proposed generalized cluster-free NOMA framework achieves the highest sum rate regardless of the variations in channel correlation.
Furthermore, CB-NOMA and enhanced CB-NOMA reduce to SDMA in the underloaded regime since there is only one user in each cluster, so they achieve the same performance.
Since BB-NOMA suffers from the \textit{SIC overuse} problem when user traffics have low channel correlations, in this case it yields worse performance than CB-NOMA/SDMA schemes.
However, when user number increases, the BB-NOMA outperforms the CB-NOMA/SDMA schemes since it can eliminate the interference among the highly channel-correlated users more adequately.
In Fig. \ref{fig_corr_ul}(b), we further compares the SIC decoding complexity, i.e., the number of matched user pairs $\sum\limits_{k\in\mathcal{K}}\sum\limits_{j\in\mathcal{K}\setminus\{k\}}\left(\alpha_{kj}+\alpha_{jk}\right)$ that implement SIC decoding in different approaches.
Owing to the cluster-free scheme, the SIC decoding complexity of the proposed framework increases with users' channel correlations, and is higher than CB-NOMA/SDMA but lower than BB-NOMA,
which demonstrates that it can achieve \textit{scenario-adaptive} SIC operations and efficient interference suppression.

\begin{figure}[!h]
	\centering
	\subfloat[Sum rate.]{\centering \scalebox{0.41}{\includegraphics{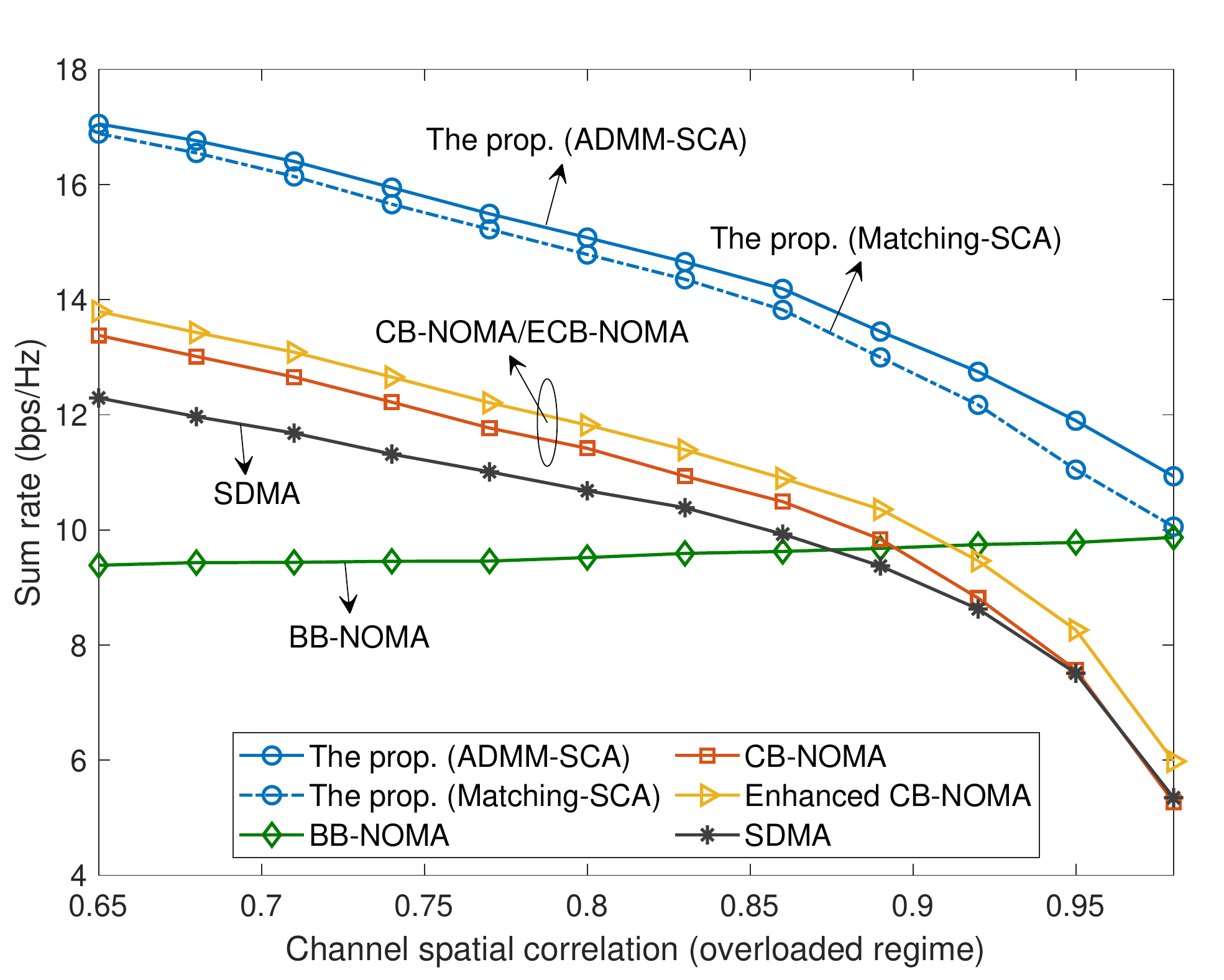}}}
	\subfloat[SIC decoding complexity.]{\centering \scalebox{0.41}{\includegraphics{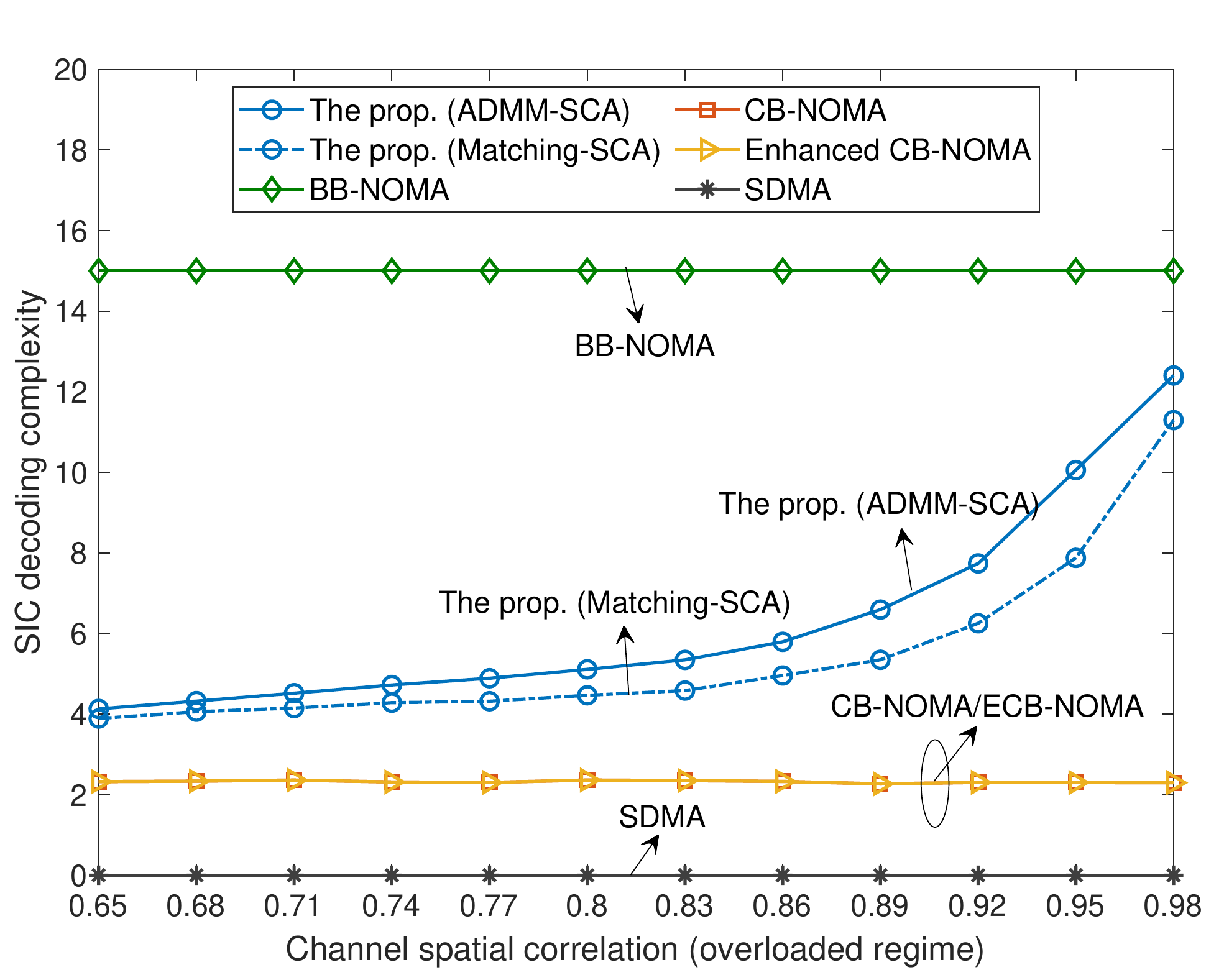}}}
    \caption{Performance comparisons under various channel correlations $corr$ in the overloaded regime. $M=4$, $K=6$.}
	\label{fig_corr_ol}
\vspace{-0.5em}
\end{figure}

Fig. \ref{fig_corr_ol} presents performance comparisons under different spatial correlations in the overloaded regime, where the number of  users is $K=6$.
From Fig. \ref{fig_corr_ol}(a), when the channel correlation increases, the sum rate increases in the BB-NOMA approach, but decreases in CB-NOMA, SDMA, and the proposed framework.
The proposed framework achieves the highest sum rate under varying channel correlations, and the performance gap is larger than that of the underloaded regime.
Furthermore, the ADMM-SCA algorithm outperforms the low-complexity Matching-SCA algorithm, and the performance gap increases with the channel correlation.
This is because the SIC decoding complexity increases with the channel correlations, which makes the control of SIC operations more complex.
Under low channel correlations, CB-NOMA and the enhanced CB-NOMA outperform the other baseline methods, while the BB-NOMA yields the worst sum rate.
Similar to the underloaded/critically loaded cases, the performance of BB-NOMA exceeds other baseline schemes when channel correlation increases.
From Fig. \ref{fig_corr_ol}(b), the proposed framework achieves the highest sum rate while maintaining a moderate SIC decoding complexity in the overloaded regime, which verifies the efficiency of the proposed framework.

\begin{figure}[!h]
\vspace{-0.5em}
  \centering
  \includegraphics[width=0.52\textwidth]{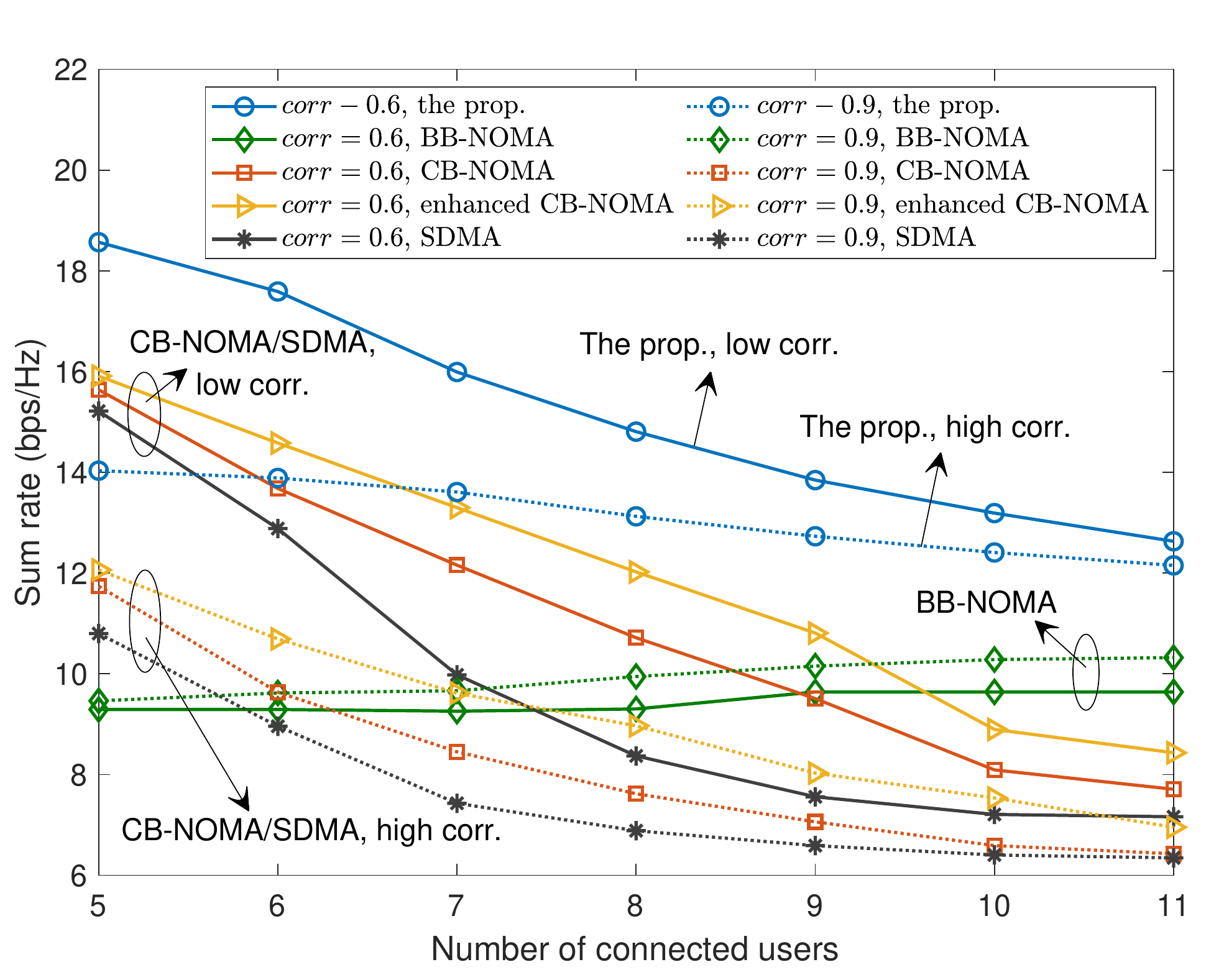}\\
  \caption{Sum rate under different numbers of connected users in the overloaded regime.}\label{fig_R_K}
\end{figure}

Fig. \ref{fig_R_K} further shows the performance comparisons under different numbers of connected users in the overloaded regime.
To reduce computational complexity, the Matching-SCA algorithm is considered.
We set $corr=0.6$ and $corr=0.9$ for low channel correlation and high channel correlation, respectively.
In both scenarios, SDMA yields the lowest sum rate than other methods.
When there are fewer users, CB-NOMA outperforms BB-NOMA and the performance gap decreases with channel correlation.
While the user number increases, BB-NOMA has higher sum rate than CB-NOMA and the performance gap increases with channel correlation.
Furthermore, the proposed framework outperforms both CB-NOMA and BB-NOMA despite the varying channel correlations owing to the efficient multiplexing.

\begin{table*}[!t]
    \begin{minipage}{1\linewidth}
        \centering
        \caption{Comparisons of Sum Rate for Different Multi-Antenna NOMA Methods}
        \label{table:comp}
        \resizebox{1\textwidth}{!}{
        \begin{tabular}{l|c|c|c|c|c}
            \hline
            \multicolumn{2}{c|}{Communication regime} & SDMA & CB-NOMA & BB-NOMA & The prop. \\ \hline
            \multirow{3}*{Low corrrelation} & Underloaded & High & High & Low & Best \\ \cline{2-6}
              & Overloaded & Medium & High & Low & Best \\ \cline{2-6}
              & Severely overloaded & Low & Medium & High & Best \\ \hline
            \multirow{3}*{High correlation} & Underloaded & High & High &Low & Best \\ \cline{2-6}
              & Overloaded & Low & Medium & High & Best \\ \cline{2-6}
              & Severely overloaded & Low & Medium & High & Best \\ \hline
        \end{tabular}
        }
	\end{minipage}
\hfill
\vspace{-0.8em}
\end{table*}

Based on the above numerical results, the performance comparisons of the proposed framework, SDMA, BB-NOMA, and CB-NOMA under different communication regimes can be summarized as Table \ref{table:comp}. It can be observed that the proposed generalized cluster-free framework leads to the highest performance regardless of the system loadings and channel correlations.

\section{Conclusions}
A novel generalized multi-antenna NOMA framework has been proposed based on the cluster-free SIC, which can reap the gains and overcome the shortcomings of traditional approaches, thus enabling a \textit{scenario-adaptive} multi-antenna NOMA paradigm for NGMA.
The transmit beamforming and the SIC operations were jointly optimized to maximize the sum rate subject to the SIC decoding conditions and data rate constraints of users.
To tackle the resulting highly-coupled NP-hard MINLP problem, an ADMM-SCA algorithm was developed to obtain the stationary solution. 
Furthermore, to accelerate the convergence and overcome the over-dependence on parameter initialization of ADMM-SCA, a Matching-SCA algorithm was proposed.
Based on an extended many-to-many matching procedure, the proposed Matching-SCA can converge to an enhanced exchange-stable matching which guarantees the local optimality.
Our numerical results showed that the proposed Matching-SCA algorithm has comparable performance to ADMM-SCA, and achieves fast convergence despite the increment of connected users.
Numerical results also verified that the proposed framework can outperform traditional multi-antenna NOMA approaches under varying channel correlations and both underloaded and overloaded regimes, which confirmed the effectiveness of the proposed framework and motivated the future research on the generalized cluster-free NOMA for empowering NGMA.

\numberwithin{equation}{section}
\section*{Appendix~A: Proposition \ref{proposition:Optimality}} \label{appendix:proof_optimality}
\renewcommand{\theequation}{A.\arabic{equation}}
\setcounter{equation}{0}
Let $\left\{\bm{\alpha}^{t*},\mathbf{W}^{t*},\mathbf{I}^{t*},\mathbf{S}^{t*},\mathbf{r}^{t*}\right\}$ and $\left\{\mathbf{W}^{t,l},\mathbf{I}^{t,l},\mathbf{S}^{t,l},\mathbf{r}^{t,l}\right\}$ denote the optimal values obtained from outer-loop iteration $t$ and the corresponding inner-loop iteration $l$ of \textbf{Algorithm \ref{alg:MatchingSCA}}, respectively.
In each inner-loop SCA iteration $l$,
functions $q_{1}\left(\mathbf{w}_{k}\right)$ and $q_2\left(I_{ik}\right)$ defined in \ref{Section_ADMM_SCA} are linearized based on the previous iterate points $\mathbf{w}_{k}^{t,l-1}$ and $I_{ik}^{t,l-1}$.
From \eqref{constraint_signalSCA}, by solving problem \eqref{PBF}, the obtained $S_{ik}^{t,l}$ and $W_{ik}^{t,l}$ always satisfy
$
S_{ik}^{t,l}\le\widehat{q}_{1}\left(\mathbf{w}_{k}^{t,l},\mathbf{w}_{k}^{t,l-1}\right)\le q_{1}\left(\mathbf{w}_{k}^{t,l}\right)=\left|\mathbf{h}_{i}^{H}\mathbf{w}_{k}^{t,l}\right|^{2},\forall i,k\in\mathcal{K}.
$
Furthermore, considering $q_{2}\left(I_{ik}^{t,l}\right)\le \widehat{q}_{2}\left(I_{i,k}^{t,l},I_{i,k}^{t,l-1}\right)$ from \eqref{Tylor_I}, the inequality
$r_{ik}^{t,l}+ q_{2}\left(I_{ik}^{t,l}\right)\le r_{ik}^{t,l}+\widehat{q}_{2}\left(I_{k}^{t,l},{I}_{k}^{t,l-1}\right)\le\log_{2}\left(I_{ik}^{t,l}+S_{ik}^{t,l}\right)
$
holds, $\forall i,k\in\mathcal{K}$.
Hence, given the fixed SIC operation variables $\bm{\alpha}^{t*}$, any feasible solution $\left\{\mathbf{W}^{t,l}, \mathbf{I}^{t,l}, \mathbf{S}^{t,l}, \mathbf{r}^{t,l} \right\}$ of problem \eqref{PBF} is also feasible to $\mathcal{P}_{1}$.
Therefore, the objective $f\left(\bm{\alpha},\mathbf{W},\mathbf{I},\mathbf{S},\mathbf{r}\right)$ is non-decreasing at each inner-loop SCA iteration $1 < l \le L^{\max}$ as \cite{SCA_converegence,SCA_convergence_2010}
\vspace{-0.5em}
\begin{equation}\label{SCAConvgence}
\begin{split}
& f\left(\bm{\alpha}^{t-1*},\mathbf{W}^{t-1*},\mathbf{I}^{t-1*},\mathbf{S}^{t-1*},\mathbf{r}^{t-1*}\right)
\le f\left(\bm{\alpha}^{t-1*},\mathbf{W}^{t,l-1},\mathbf{I}^{t,l-1},\mathbf{S}^{t,l-1},\mathbf{r}^{t,l-1}\right)
\\
\le & f\left(\bm{\alpha}^{t-1*},\mathbf{W}^{t,l},\mathbf{I}^{t,l},\mathbf{S}^{t,l},\mathbf{r}^{t,l}\right)
\le f\left(\bm{\alpha}^{t-1*},\mathbf{W}^{t,L^{\max}},\mathbf{I}^{t,L^{\max}},\mathbf{S}^{t,L^{\max}},\mathbf{r}^{t,L^{\max}}\right).
\end{split}
\end{equation}
\vspace{-0.5cm}

Moreover, according to \textbf{Proposition \ref{proposition:Convergence}}, the sum rate is non-decreasing after each enhanced swap operation.
Thus, the sequence $\left\{f\left(\bm{\alpha}^{t*},\mathbf{W}^{t*},\mathbf{I}^{t*},\mathbf{S}^{t*},\mathbf{r}^{t*}\right)\right\}_{t\in\{1,2,...,T_{\mathrm{MSCA}}^{\max}\}}$
has monotonic convergence.
Considering the limited transmit power, \textbf{Algorithm \ref{alg:MatchingSCA}} will terminate until there are no SCA updates nor swap operations can further increase the sum srate. Hence, the local optimality of the solutions can be guaranteed, which completes the proof.
\vspace{-0.3cm}

\ifCLASSOPTIONcaptionsoff
  \newpage
\fi


\begin{thebibliography}{1}

\begin{spacing}{1.12}
\bibitem{Cisco}
U. Cisco, ``Cisco annual internet report (2018-2023) white paper,'' 2020.

\bibitem{6GIoE}
W. Saad, M. Bennis and M. Chen, ``A vision of 6G wireless systems: applications, trends, technologies, and open research problems,'' \textit{IEEE Netw.}, vol. 34, no. 3, pp. 134-142, May/Jun. 2020.

\bibitem{6GRoadmap}
K. B. Letaief, W. Chen, Y. Shi, J. Zhang, and Y. A. Zhang, ``The roadmap to 6G: AI empowered wireless networks,'' \textit{IEEE Commun. Mag.}, vol. 57, no. 8, pp. 84-90, Aug. 2019.



\bibitem{EvolNOMA}
Y. Liu, S. Zhang, X. Mu, Z. Ding, R. Schober, N. Al-Dhahir, E. Hossain, and X. Shen,
``Evolution of NOMA toward next generation multiple access (NGMA) for 6G,'' \textit{IEEE J. Sel. Areas Commun.}, early access, 2022.


\bibitem{PDNOMA_2017}
S. M. R. Islam, N. Avazov, O. A. Dobre and K. -s. Kwak, ``Power-domain non-orthogonal multiple access (NOMA) in 5G systems: potentials and challenges,'' 
\textit{IEEE Commun. Surv. Tut.}, vol. 19, no. 2, pp. 721-742, Secondquarter 2017.

\bibitem{NOMA5G_2017}
Y. Liu, Z. Qin, M. Elkashlan, Z. Ding, A. Nallanathan and L. Hanzo, ``Nonorthogonal multiple access for 5G and beyond,'' \textit{Proc. IEEE}, vol. 105, no. 12, pp. 2347-2381, Dec. 2017.

\bibitem{Fairness_NOMA_2018}
S. M. R. Islam, M. Zeng, O. A. Dobre and K. -S. Kwak, ``Resource allocation for downlink NOMA systems: Key techniques and open issues,'' 
\textit{IEEE Wireless Commun.}, vol. 25, no. 2, pp. 40-47, Apr. 2018.

\bibitem{MIMONOMA_Huang_2018}
Y. Huang, C. Zhang, J. Wang, Y. Jing, L. Yang and X. You, ``Signal processing for MIMO-NOMA: Present and future challenges,'' 
\textit{IEEE Wireless Commun.}, vol. 25, no. 2, pp. 32-38, Apr. 2018.

\bibitem{MassiveAccess}
X. Chen, D. W. K. Ng, W. Yu, E. G. Larsson, N. Al-Dhahir and R. Schober, ``Massive access for 5G and beyond,'' \textit{IEEE J. Sel. Areas Commun.}, vol. 39, no. 3, pp. 615-637, Mar. 2021.

\bibitem{MIMONOMA_2016}
Z. Ding, F. Adachi and H. V. Poor, ``The application of MIMO to non-orthogonal multiple access,'' \textit{IEEE Trans. Wireless Commun.}, vol. 15, no. 1, pp. 537-552, Jan. 2016.


\bibitem{MIMONOMA_2018}
Y. Liu, H. Xing, C. Pan, A. Nallanathan, M. Elkashlan, and L. Hanzo, ``Multiple-antenna-assisted non-orthogonal multiple access,'' \textit{IEEE Wireless Commun.}, vol. 25, no. 2, pp. 17-23, Apr. 2018.


\bibitem{BBNOMA_2016_Hanif}
M. F. Hanif, Z. Ding, T. Ratnarajah, and G. K. Karagiannidis, ``A minorization-maximization method for optimizing sum
rate in the downlink of non-orthogonal multiple access systems,'' \textit{IEEE Trans. Signal Process.}, vol. 64, no. 1, pp. 76-88, Jan. 2016.


\bibitem{BBNOMA_2015}
Q. Sun, S. Han, Z. Xu, S. Wang, I. Chih-Lin, and Z. Pan, ``Sum rate optimization for MIMO non-orthogonal multiple access systems,'' \textit{Proc. IEEE Wireless Commun. Netw. Conf.
(IEEE WCNC)}, New Orleans, LA, USA, Mar. 2015, pp. 747-752.

\bibitem{BBNOMA_2016_Chen}
Z. Chen, Z. Ding, X. Dai, and G. K. Karagiannidis, ``On the application of quasi-degradation to MISO-NOMA downlink,''
\textit{IEEE Trans. Signal Process.}, vol. 64, no. 23, pp. 6174-6189, Dec. 2016.


\bibitem{BBNOMA_Chen}
C. Chen, W. Cai, X. Cheng, L. Yang and Y. Jin, ``Low complexity beamforming and user selection schemes for 5G MIMO-NOMA systems,'' \textit{IEEE J. Sel. Areas Commun.}, vol. 35, no. 12, pp. 2708-2722, Dec. 2017.


\bibitem{CBNOMA_OMA}
M. Zeng, A. Yadav, O. A. Dobre, G. I. Tsiropoulos and H. V. Poor, ``Capacity comparison between MIMO-NOMA and MIMO-OMA with multiple users in a cluster,'' \textit{IEEE J. Sel. Areas Commun.}, vol. 35, no. 10, pp. 2413-2424, Oct. 2017.

\bibitem{EE_CBNOMA_2019}
M. Zeng, W. Hao, O. A. Dobre and H. V. Poor, ``Energy-efficient power allocation in uplink mmWave massive MIMO with NOMA,'' 
\textit{IEEE Trans. Veh. Tech.}, vol. 68, no. 3, pp. 3000-3004, Mar. 2019.


\bibitem{CBNOMA_2019_Hu}
X. Hu, C. Zhong, X. Chen, W. Xu and Z. Zhang, ``Cluster grouping and power control for angle-domain mmWave MIMO NOMA systems,'' 
\textit{IEEE J. Sel. Topics Signal Process.}, vol. 13, no. 5, pp. 1167-1180, Sept. 2019

\bibitem{Multicell_CBNOMA_2020}
Y. Fu, M. Zhang, L. Salaün, C. W. Sung and C. S. Chen, ``Zero-forcing oriented power minimization for multi-cell MISO-NOMA systems: A joint user grouping, beamforming, and power control perspective,'' \textit{IEEE J. Sel. Areas Commun.}, vol. 38, no. 8, pp. 1925-1940, Aug. 2020.

\bibitem{NOMABFDesign_Ding}
Z. Ding, ``NOMA beamforming in SDMA networks: Riding on existing beams or forming new ones?'' \textit{IEEE Commun. Lett.}, early access, 2022.

\bibitem{ConvexOpt}
S. Boyd and L. Vandenberghe, \textit{Convex Optimization. Cambridge}, U.K.: Cambridge Univ. Press, 2004.

\bibitem{ADMM_1976}
D. Gabay and B. Mercier, ``A dual algorithm for the solution of nonlinear variational problems via finite element approximation,''
\textit{Comput. Math. Appl.}, vol. 2, no. 1, pp. 17-40, 1976.

\bibitem{ADMM_Boyd}
S. Boyd, N. Parikh, E. Chu, B. Peleato and J. Eckstein, ``Distributed optimization and statistical learning via the alternating direction method of multipliers,'' \textit{Found. Trends Mach. Learn.}, vol. 3, no. 1, pp. 1-122, Jan. 2011.

\bibitem{SCA_1978}
B. R. Marks and G. P. Wright, ``A general inner approximation algorithm for nonconvex mathematical programs,''
\textit{Oper. Res.}, vol. 26, no. 4, pp. 681-683, 1978.


\bibitem{CVX}
M. Grant and S. Boyd, ``CVX: MATLAB software for disciplined convex programming,'' 2016, [online] Available: http://cvxr.com/cvx.


\bibitem{SCA_converegence}
G. R. Lanckriet and B. K. Sriperumbudur, ``On the convergence of the concave-convex procedure,''
in \textit{Proc. Adv. Neural Inf. Process. Syst.}, 2009, pp. 1759-1767.


\bibitem{ADMM_convergence}
G. Li and T. K. Pong, ``Global convergence of splitting methods for nonconvex composite optimization,''
\textit{SIAM J. Optimization}, vol. 25, no. 4, pp. 2434-2460, 2015.


\bibitem{MatchingExternality_Zhao}
J. Zhao, Y. Liu, K. K. Chai, Y. Chen and M. Elkashlan, ``Many-to-Many matching with externalities for device-to-device communications,''
\textit{IEEE Wireless Commun. Lett.}, vol. 6, no. 1, pp. 138-141, Feb. 2017.

\bibitem{MatchingExternality_Ni}
W. Ni, X. Liu, Y. Liu, H. Tian and Y. Chen, ``Resource allocation for multi-cell IRS-aided NOMA networks,'' \textit{IEEE Trans. Wireless Commun.}, vol. 20, no. 7, pp. 4253-4268, Jul. 2021.

\bibitem{Stability_Bodine}
E. Bodine-Baron \textit{et al.}, ``Peer effects and stability in matching markets,'' \textit{International Symposium on Algorithmic Game Theory}, 2011.

\bibitem{ChannelCorr}
J. P. Kermoal, L. Schumacher, K. I. Pedersen, P. E. Mogensen and F. Frederiksen, ``A stochastic MIMO radio channel model with experimental validation,''
\textit{IEEE J. Sel. Areas Commun.}, vol. 20, no. 6, pp. 1211-1226, Aug. 2002.

\bibitem{UserClustering_Dai}
L. Dai, B. Wang, M. Peng and S. Chen, ``Hybrid precoding-based millimeter-wave massive MIMO-NOMA with simultaneous wireless information and power transfer,''
\textit{IEEE J. Sel. Areas Commun.}, vol. 37, no. 1, pp. 131-141, Jan. 2019.

\bibitem{SCA_convergence_2010}
A. Beck, A. Ben-Tal, and L. Tetruashvili, ``A sequential parametric convex approximation method with applications to nonconvex truss
topology design problems,'' \textit{J. Global Optim.}, vol. 47, no. 1, pp. 29-51, 2010.
\end{spacing}
\end{thebibliography}
\end{document}